\documentclass[12pt]{article}
\def\cdate{{February 9, 2009}}
\usepackage{amsthm,amscd}
\usepackage{mathrsfs}
\usepackage{verbatim}
\usepackage{amsfonts,amsmath,latexsym,amssymb,txfonts,bm}
\usepackage[american]{babel}

\makeatletter
\renewcommand{\@evenhead}{\raisebox{0pt}[\headheight][0pt]{\vbox{\hbox
to \textwidth{\thepage\hfil\strut\textit{\leftmark}}\hrule}}}
\renewcommand{\@oddhead}{\raisebox{0pt}[\headheight][0pt]{\vbox{\hbox
to \textwidth{\textit{\rightmark}\hfil\strut\thepage}\hrule}}}

\makeatother

\def\II{{\mathbb I}}

\def\tr{\mathrm{ tr\,}}
\def\Tr{\mathrm{ Tr\,}}

\def\Det{\mathrm{ Det\,}}

\def\Res{\mathrm{ Res\,}}

\def\be{\begin{equation}}
\def\ee{\end{equation}}
\def\bea{\begin{eqnarray}}
\def\eea{\end{eqnarray}}
\def\bed{\begin{definition}{\ }}
\def\eed{\end{definition}}
\def\bd{\begin{description}}
\def\ed{\end{description}}
\def\bc{\begin{center}}
\def\ec{\end{center}}

\newtheorem{definition}{Definition}

\def\sideremark#1{\ifvmode\leavevmode\fi\vadjust{\vbox to0pt{\vss
\hbox to 0pt{\hskip\hsize\hskip1em
\vbox{\hsize2cm\tiny\raggedright\pretolerance10000
\noindent #1\hfill}\hss}\vbox to8pt{\vfil}\vss}}}

\title{}
\author{}

\begin{document}

\begin{titlepage}
\thispagestyle{empty}
\null
\vspace{-3cm}
\hspace*{50truemm}{\hrulefill}\par\vskip-4truemm\par
\hspace*{50truemm}{\hrulefill}\par\vskip5mm\par
\hspace*{50truemm}{{\large\sc New Mexico Tech {\rm
(\cdate)
}}}
\vskip4mm\par
\hspace*{50truemm}{\hrulefill}\par\vskip-4truemm\par
\hspace*{50truemm}{\hrulefill}
\par
\bigskip
\bigskip
\vfill
\centerline{\huge\bf Low-Energy Effective Action}
\bigskip
\centerline{\huge\bf in Non-Perturbative Electrodynamics}
\bigskip
\centerline{\huge\bf in Curved Spacetime}
\bigskip
\bigskip
\centerline{\Large\bf Ivan G. Avramidi and Guglielmo Fucci}
\bigskip
\centerline{\it New Mexico Institute of Mining and Technology}
\centerline{\it Socorro, NM 87801, USA}
\centerline{\it E-mail: iavramid@nmt.edu, gfucci@nmt.edu}
\bigskip
\medskip
\vfill

{\narrower
\par
We study the heat kernel for the Laplace type partial differential operator
acting on smooth sections of a complex spin-tensor bundle over a generic
$n$-dimensional Riemannian manifold. Assuming that the curvature of the $U(1)$
connection (that we call the electromagnetic field) is constant we compute the
first two coefficients of the non-perturbative asymptotic expansion of the heat
kernel which are of zero and the first order in Riemannian curvature and of
arbitrary order in the electromagnetic field. We apply these results to the
study of the effective action in non-perturbative electrodynamics in four
dimensions and derive a generalization of the Schwinger's result for the
creation of scalar and spinor particles in electromagnetic field induced by the
gravitational field. We discover a new infrared divergence in the imaginary
part of the effective action due to the gravitational corrections, which seems
to be a new physical effect.

\par}


\end{titlepage}


\section{Introduction}
\setcounter{equation}0

The effective action is one of the most powerful tools in quantum field theory
and quantum gravity (see
\cite{schwinger51,dewitt03,avramidi00,avramidi02,avramidi09}). The effective
action is a functional of the background fields that encodes, in principle, all
the information of quantum field theory. It determines the full one-point
propagator and the full vertex functions and, hence, the whole $S$-matrix.
Moreover, the variation of the effective action gives the effective equations
for the background fields, which makes it possible to study the back-reaction
of quantum processes on the classical background. In particular, the low energy
effective action (or the effective potential) is the most appropriate tool for
investigating the structure of the physical vacuum in quantum field theory.

The effective action is expressed in terms of the propagators and the vertex
functions. One of the most powerful methods to study the propagators is the
heat kernel method, which was originally proposed by Fock \cite{fock37} and
later generalized by Schwinger \cite{schwinger51} who also applied it to the
calculation of the one-loop  effective action in quantum electrodynamics.
Finally, De Witt reformulated it in the geometrical language and applied it to
the case of gravitational field (see his latest book \cite{dewitt03}).

In particular Schwinger solved exactly the case of a constant electromagnetic
field and derived an heat kernel integral representation for the effective
action. He showed that the heat kernel becomes a meromorphic function and a
careful evaluation of the integral leads to an imaginary part of the effective
action. Schwinger computed the imaginary part of the effective action and
showed that it describes the effect of creation of electron-positron pairs by
the electric field. This effect is now called the Schwinger mechanism. This is
an essentially non-perturbative effect (non-analytic in electric field) that
vanishes exponentially for weak electric fields.

Therefore its evaluation requires non-perturbative techniques for the
calculation of the heat kernel in the situation when curvatures (but not their
derivatives) are large (low energy approximation). A powerful approach to the
calculation of the low-energy heat kernel expansion was developed in
non-Abelian gauge theories and quantum gravity in
\cite{avramidi93,avramidi94a,avramidi95,avramidi95a,avramidi96,
avramidi08b,avramidi08,avramidi08a}. While the papers
\cite{avramidi93,avramidi95,avramidi95a} dealt with the constant
electromagnetic field in flat space, the papers
\cite{avramidi94a,avramidi96,avramidi08b} dealt with symmetric spaces (pure
gravitational field in absence of an electromagnetic field). The difficulty of
combining the gauge fields and gravity was finally overcome in the papers
\cite{avramidi08,avramidi08a}, where homogeneous bundles with parallel
curvature on symmetric spaces was studied.

In \cite{avramidi08b} we computed the heat kernel for the covariant Laplacian
with a strong covariantly constant electromagnetic field in an arbitrary
gravitational field. We evaluated the first three coefficients of the heat
kernel asymptotic expansion in powers of Riemann curvature $R$ but \emph{in all
orders} of the electromagnetic field $F$. This is equivalent to a partial
summation in the heat kernel asymptotic expansion as $t\to 0$ of all powers of
$F$ in terms which are linear and quadratic in Riemann curvature $R$. In the
present paper we use those results to compute {\it explicitly} the terms linear
in the Riemann curvature in the non-perturbative heat kernel expansion for the
scalar and the spinor fields and compute their contribution to the imaginary
part of the effective action. In other words, we generalize the Schwinger
mechanism to the case of a strong electromagnetic field in a gravitational
field and compute the {\it gravitational corrections to the original Schwinger
result}.

\section{Setup of the Problem}
\setcounter{equation}0

Let $M$ be a $n$-dimensional compact Riemannian manifold
(with positive-definite metric $g_{\mu\nu}$) without boundary and
${\cal S}$ be a complex spin-tensor vector bundle over $M$ realizing a
representation of
the group ${\rm Spin}(n)\otimes U(1)$.
Let $\varphi$ be a section of the bundle ${\cal S}$
and $\nabla$ be the total connection on the bundle $\mathcal{S}$ (including the
spin connection as well as the ${\rm U}(1)$-connection). Then the commutator of
covariant derivatives defines the curvatures
\begin{equation}\label{0}
[\nabla_\mu,\nabla_\nu]\varphi=({\cal R}_{\mu\nu}+i F_{\mu\nu})\varphi\;,
\end{equation}
where $F_{\mu\nu}$ is the curvature of the $U(1)$-connection (which will be
also
called the electromagnetic field) and
${\cal R}_{\mu\nu}$ is the curvature of the spin connection defined by
\be
{\cal R}_{\mu\nu}=\frac{1}{2}R^{ab}{}_{\mu\nu}\Sigma_{ab}\,,
\ee
with $\Sigma_{ab}$ being the generators of the spin group ${\rm Spin}(n)$
satisfying
the commutation relations
\be
[\Sigma_{ab},\Sigma^{cd}]
=4\delta^{[c}{}_{[a}\Sigma^{d]}{}_{b]}\,.
\label{280}
\ee
Note that for the scalar fields ${\cal R}_{\mu\nu}=0$ and for the spinor fields
\be
\Sigma_{ab}=\frac{1}{2}\gamma_{ab}\,,
\ee
where $\gamma_{ab}=\gamma_{[a}\gamma_{b]}$
(more generally, we define
$\gamma_{a_1\dots a_m}=\gamma_{[a_1}\cdots\gamma_{a_m]}$)
and $\gamma_a$ are
the Dirac matrices
generating the Clifford algebra
\be
\gamma_a\gamma_b+\gamma_b\gamma_a=2g_{ab}\II\,.
\ee

\subsection{Differential Operators}

In the present paper we consider a second-order Laplace type partial
differential operator,
\be
\label{1}
L=-\Delta+\xi R + Q,
\ee
where $\Delta=g^{\mu\nu}\nabla_\mu\nabla_\nu$ is the Laplacian,
$\xi$ is a constant parameter,
and $Q$ is a smooth endomorphism of the bundle ${\cal S}$.
This operator is elliptic and self-adjoint and has a
positive-definite leading symbol.
Usually, for scalar fields we set
\be
Q^{\rm scalar}=0\,.
\ee
Moreover, for canonical scalar fields the coupling
\be
\xi^{\rm scalar}=
\left\{
\begin{array}{ll}
0 & \mbox{for canonical scalar fields}\,,\\[10pt]
\displaystyle
\frac{(n-2)}{4(n-1)} & \mbox{for conformal scalar fields}\,.
\end{array}
\right.
\ee
Another important case is the square of the Dirac operator acting on
spinor fields
\be
L=D^2\,,
\ee
where
\be
D=i\gamma^\mu\nabla_\mu\,.
\ee
It is easy to see that in this case
we have
\be
\xi^{\rm spinor}=\frac{1}{4}
\ee
and
\be
Q^{\rm spinor}=-\frac{1}{2}iF_{\mu\nu}\gamma^{\mu\nu}\,.
\ee

\subsection{Effective Action}

The object of primary interest in quantum field theory is the
(Euclidean)
one-loop effective action determined by the formal
determinant
\be
\Gamma_{(1)}=\sigma\log\Det(L+m^2)\,,
\ee
where $\sigma$ is the fermion number of the field equal to $(+1)$ for
boson fields and $(-1)$ for fermion fields,
$m$ is a
mass parameter, which is assumed to be sufficiently
large so that the operator $(L+m^2)$ is positive.
Notice that the usual factor $\frac{1}{2}$ is missing because the field
is complex, which is equivalent to the contribution of two real fields.
Of course, this formal
expression is divergent.
To rigorously define the determinant of a differential operator
one needs to introduce some regularization and then to renormalize it.
One of the best ways to do it is via the heat kernel method.

In a physical theory the effective action describes the
in-out vacuum transition amplitude via
\be
\left<{\rm out}|{\rm in}\right>=\exp[i\Gamma_{(1)}]\,.
\ee
The real part of the effective
action
describes the polarization of vacuum of quantum fields by the background
fields and the imaginary part describes
the creation of particles. Namely,
the probability of particles production (in the whole spacetime)
is given by
\be
P=1-\left|\left<{\rm out}|{\rm in}\right>\right|^2
=1-\exp\left[-2\,{\rm Im}\,\Gamma_{(1)}\right]\,.
\ee
Of course, the unitarity requires that the imaginary part of the effective
action should be positive
\be
{\rm Im}\,\Gamma_{(1)}\ge 0\,.
\ee
Notice that usually, when the imaginary part of the effective action
is small, we just have
\be
P\approx 2\,{\rm Im}\,\Gamma_{(1)}\,.
\ee
The one-loop effective Lagrangian is defined by
\be
\Gamma_{(1)}=\int\limits\limits_M dx\;g^{1/2}{\cal L}\,.
\ee
Therefore, the rate of the particles production per unit volume per
unit time is given by the imaginary part of the effective Lagrangian
\be
R=\frac{P}{VT}
\approx 2\,{\rm Im}\,{\cal L}\,.
\ee

\subsection{Spectral Functions}

The heat kernel for the operator ${L}$ is defined as
the solution of the heat equation
\begin{equation}
\left(\partial_{t}+{L}\right)U(t|x,x^{\prime})=0\;,
\end{equation}
with the initial condition
\begin{equation}
U(0|x,x^{\prime})=\delta(x,x^{\prime})\;.
\end{equation}
where $\delta(x,x^{\prime})$ is the covariant scalar delta function
on the bundle ${\cal S}$.
The heat kernel diagonal is defined by
\be
U^{\rm diag}(t)=U(t|x,x)\,.
\ee

One of the best ways to describe the
spectral properties of the operator ${L}$ is via the heat trace
\be
\textrm{Tr}\;\exp(-t{L})=
\int\limits_{M}dx\;g^{1/2}\Theta(t)\,,
\ee
where $dx$ is the Lebesgue measure on the manifold $M$, $g=\det g_{\mu\nu}$
and
\be
\Theta(t)=\textrm{tr}\;U^{\rm diag}(t)\,.
\ee
Here $\textrm{tr}$ denotes the fiber trace over the bundle ${\cal S}$.

The determinant of the operator can be defined within the so-called
zeta-function regularization as follows.
First, one defines the zeta function by
\be
\zeta(s)=\mu^{2s}\Tr \left(L+m^2\right)^{-s}
=\int\limits_M dx\;g^{1/2}Z(s)\,,
\ee
where
\be
Z(s)=\frac{\mu^{2s}}{\Gamma(s)}\int\limits_0^\infty
dt\; t^{s-1} e^{-tm^2}\Theta(t)\,,
\ee
and $\mu$ is a renormalization parameter
introduced to preserve dimensions. The zeta function $\zeta(s)$ is a
meromorphic
function of $s$ analytic at $s=0$. This enables one to define
the (zeta-regularized) functional determinant of the operator $(L+m^2)$
by
\be
\Det(L+m^2)=\exp[-\zeta'(0)]\,,
\ee
where $\zeta'(s)=\frac{d}{ds}\zeta(s)$.
Therefore, the one-loop effective action is simply
\be
\Gamma_{(1)}=-\sigma\zeta'(0)\,,
\ee
and the one-loop effective Lagrangian is given by
\be
{\cal L}=-\sigma Z'(0)\,.
\ee
The effective Lagrangian can be also defined simply in the cut-off
regularization by
\be
{\cal L}=-\sigma\int\limits_{\varepsilon\mu^2}^\infty
\frac{dt}{t}e^{-tm^2}\Theta(t)\,,
\ee
where $\varepsilon$ is a regularization parameter, which should be
set to zero after subtracting the divergent terms.
Another regularization is the dimensional regularization, in which one
simply defines the effective action by the formal integral
\be
{\cal L}=-\sigma\mu^{2\varepsilon}\int\limits_{0}^\infty
\frac{dt}{t}e^{-tm^2}\Theta(t)\,,
\ee
where the heat trace is formally computed in complex dimension
$(n-2\varepsilon)$
with sufficiently large real part of $\varepsilon$ so that the integral is
finite. The renormalized effective action is obtained then by
subtracting the simple pole in $\varepsilon$.

For elliptic operators (in the Euclidean setup) the heat trace is a smooth
function of $t$; in many cases it is even an analytic function of $t$ in the
neighborhood of the positive real axis. However, in the physical case for
hyperbolic operators (in the Lorentzian setup) the heat trace can have
singularities even on the positive real axis of $t$. As we will show later
in the approximation under consideration (for constant electromagnetic field)
it becomes a meromorphic
function of $t$ with an essential singularity at $t=0$ and some poles
$t_k$, $k=1,2,\dots,$,
on the positive real axis.
It turns out that the imaginary part of the effective action does not depend
on the regularization method and is uniquely defined by the contribution
of these poles. These poles should be avoided from above, which gives
\be
{\rm Im}\;{\cal L}
=-\sigma \pi\sum_{k=1}^\infty
\Res\left\{t^{-1} e^{-t m^2} \Theta(t); t_k\right\}\,.
\label{232xxx}
\ee
This method was first elaborated and used by Schwinger
\cite{schwinger51} in quantum electrodynamics to calculate the
electron-positron pair production by a constant electric field.
One of the goal of our work is to {\it generalize the Schwinger results}
for the case of constant electromagnetic field in a gravitational field.
We will compute the {\it extra contribution} to the particle production by
a constant electromagnetic field {\it induced by
the gravitational field}.

\subsection{Heat Kernel Asymptotic Expansion}

It is well known \cite{gilkey95} that the heat kernel diagonal has the
asymptotic expansion as $t\rightarrow 0$ (see also
\cite{avramidi91,avramidi99,avramidi00,vassilevich03})
\begin{equation}
U^{\rm diag}(t)\sim(4\pi t)^{-n/2}
\sum_{k=0}^{\infty}t^{k}a_{k}\;,
\end{equation}
where $a_{k}$ are the local heat kernel coefficients.
Then the trace of the diagonal heat kernel
has the corresponding asymptotic expansion
\begin{equation}
\Theta(t)
\sim(4\pi t)^{-n/2}\sum_{k=0}^{\infty}t^{k}A_{k}\;,
\end{equation}
where
\begin{equation}
A_{k}=\textrm{tr}\;a^{\rm }_{k}\,.
\end{equation}

The diagonal heat kernel coefficients $a^{\rm }_k$ are polynomials in the
jets of the metric, the $U(1)$-connection
and the potential term $Q$, that is, in
the curvature tensors, $Q$, and their derivatives.
The lower order diagonal heat kernel coefficients are well known
\cite{gilkey95,avramidi91,avramidi00}
\bea
a^{\rm }_0&=&1\,,
\\[10pt]
a^{\rm }_1&=&-Q+\left(\frac{1}{6}-\xi\right)R\,.
\label{120iga}
\eea
To avoid confusion we should stress that the normalization of the coefficients
$a_k$ differs from the papers \cite{avramidi91,avramidi99,avramidi00}.

In our previous paper \cite{avramidi08b}
we studied the case of a {\it parallel $U(1)$ curvature}
(covariantly constant electromagnetic field), i.e.
\begin{equation}
\label{2x}
\nabla_{\mu}F_{\alpha\beta}=0\;.
\end{equation}
In the present paper we will also assume that the potential term $Q$
is covariantly constant too
\be
\nabla_\mu Q=0\,.
\ee
By summing up all powers of $F$ in the asymptotic expansion
of the heat kernel diagonal we obtained a {\it new (non-perturbative)
asymptotic expansion}
\be
U^{\rm diag}(t)\sim(4\pi t)^{-n/2}\exp\left(-tQ\right)J(t)
\sum_{k=0}^{\infty}t^{k} b_{k}(t)\;,
\label{222zza}
\ee
where
\be
J(t)=\det\left(\frac{tiF}{\sinh(tiF)}\right)^{1/2}
\label{25zz}
\ee
and $b_k(t)$ are the modified heat kernel coefficients
which are analytic functions of $t$ at $t=0$ which depend on $F$ only in the
dimensionless combination $tF$.
Here and everywhere below all
functions of the $2$-form $F$ are analytic at $0$ and should be understood in
terms of a power series in the matrix $F=(F^\mu{}_\nu)$. Notice the
{\it position of indices} here, it is important! There is a difference here
between Euclidean case and the
Lorentzian one since the raising of indices by a Minkowski metric does change
the properties of the matrix $F$.
Also, here $\det$ denotes the determinant with respect to the tangent space
indices.

The fiber trace of the heat kernel diagonal
has then the asymptotic expansion
\begin{equation}
\Theta(t)
\sim(4\pi t)^{-n/2}\Phi(t)\sum_{k=0}^{\infty}t^{k}
B_{k}(t)\;,
\label{225xx}
\end{equation}
where
\be
\Phi(t)=J(t)\textrm{tr}\, \exp\left(-tQ\right)\,,
\ee
\be
B_k(t)=\frac{\textrm{tr}\, \exp\left(-tQ\right)b_{k}(t)}
{\textrm{tr}\, \exp\left(-tQ\right)}\;
\ee
are {\it new (non-perturbative) heat kernel coefficients}
of the operator $L$. The integrals $\int_M dx g^{1/2}B_k(t)$ are then
the spectral invariants of the operator $L$.

This expansion can be described more rigorously as follows. We rescale the
$U(1)$-curvature $F$ by
\be
F \mapsto F(t)=t^{-1}\tilde F\,,
\ee
so that $tF(t)=\tilde F$ is independent of $t$. Then the operator
${L}(t)$ becomes dependent on $t$ (in a singular way!). However, the
heat trace still has a nice asymptotic expansion (\ref{225xx})
as $t\to 0$,
where the coefficients $B_k$ are expressed in terms of $\tilde F=tF(t)$,
and, therefore, are independent of $t$. Thus, what we are doing is the {\it
asymptotic expansion of the heat trace for a particular case of a singular (as
$t\to 0$) time-dependent operator ${L}(t)$}.

\section{Calculation of the Coefficient $B_1(t)$ }
\setcounter{equation}0

For the first two coefficients we obtained
\cite{avramidi08b}
\bea
b_0^{\rm }(t)&=&1\,,
\\
\label{70}
b^{\rm }_{1}(t)&=&
\left\{\Sigma_{\mu\alpha}W_{\nu\beta}(t)
+V_{\mu\alpha\nu\beta}(t)\right\}
R^{\mu\alpha\nu\beta}
\eea
where
\bea
W(t)
&=&\frac{1}{2}\left(\coth(tiF)-\frac{1}{tiF}\right)\;,
\\
\label{72c}
V^{\mu\alpha}{}_{\nu\beta}(t)
&=&
\left(\frac{1}{3}-\xi\right)\delta^{[\mu}_\nu\delta^{\alpha]}_\beta
+\int\limits_{0}^{1}d\tau
\Bigg\{
-\frac{1}{24}\mathcal{B}^{[\mu}{}_{[\nu}(\tau)
{\cal Z}^{\alpha]}{}_{\beta]}(\tau)
+\frac{1}{6}\mathcal{A}^{[\mu\alpha]}(\tau)\mathcal{A}_{[\nu\beta]}(\tau)
\nonumber\\
&&
-\frac{1}{12}\mathcal{A}^{[\mu}{}_{[\nu}(\tau)
\mathcal{A}^{\alpha]}{}_{\beta]}(\tau)
-\frac{1}{4}\mathcal{A}^{[\mu}{}_{[\nu}(\tau)
\mathcal{A}_{\beta]}{}^{\alpha]}(\tau)
\Bigg\} \;,
\eea
and
\bea
\label{71}
\mathcal{A}(\tau)
&=&
\frac{1}{2}\frac{\exp[(1-2\tau)tiF]-\exp(-tiF)}{\sinh(tiF)}\;,
\\[5pt]
\label{71b}
\mathcal{B}(\tau)
&=&
\frac{\coth(tiF)}{tiF}
-\frac{1}{tiF\sinh(tiF)}\cosh[(1-2\tau)tiF]\;,
\\[5pt]
{\cal Z}(\tau)
&=&
3tiF\coth(tiF)
+\frac{tiF}{\sinh(tiF)}\cosh[(1-2\tau)tiF]\;.
\eea
The trace coefficients are then given by
\bea
B_0(t)&=&1\,,
\\
\label{70b}
B_{1}(t)
&=&
\bigg\{\Psi_{\mu\alpha}(t)W_{\nu\beta}(t)
+V_{\mu\alpha\nu\beta}(t)\bigg\}
R^{\mu\alpha\nu\beta}
\;,
\eea
where
\bea
\Psi(t)_{\mu\alpha}
&=&\;\frac{\textrm{tr}\, \exp\left(-tQ\right)\Sigma_{\mu\alpha}}
{\textrm{tr}\, \exp\left(-tQ\right)}\,.
\eea

\subsection{Spectral Decomposition}

To evaluate it we use the spectral
decomposition of the matrix $F=(F^\mu{}_\nu)$,
\be
F=\sum_{k=1}^{N} B_k E_k\,,
\ee
where $B_k$ are some real
invariants and $E_k=(E_k{}^\mu{}_\nu)$ are some
matrices
satisfying the equations
\be
E_k{}_{\mu\nu}=-E_k{}_{\nu\mu}\,,
\ee
\be
E^k_{\mu[\nu}E^k_{\alpha\beta]}=0
\,,
\ee
and for $k\ne m$
\be
E_k E_{m}=0\,.
\ee
Here, of course, $N\le [n/2]$.
The invariants $B_k$ (that we call ``magnetic fields'') should not be confused
with the heat trace coefficients
$B_0$ and $B_1$.

Next, we define the
matrices $\Pi_k=(\Pi_k{}^\mu{}_\nu)$ by
\be
\Pi_k=-E_k^2\,.
\ee
They satisfy the equations
\be
\Pi_k{}_{\mu\nu}=\Pi_k{}_{\nu\mu}\,,
\ee
\be
E_k\Pi_k=\Pi_kE_k=E_k\,,
\ee
and for $k\ne m$
\be
E_k \Pi_m=\Pi_m E_k=0\,, \qquad
\Pi_k\Pi_m=0\,.
\ee

To compute functions of the matrix $F$ we need to know its eigenvalues.
We distinguish two different cases.

\paragraph{Euclidean Case.} In this case the metric has Euclidean
signature $(++\cdots+)$ and the non-zero
eigenvalues of the matrix $F$ are
$\pm iB_1$, \dots, $\pm iB_N$, (which are all imaginary).
Of course, it may also have a number of zero eigenvalues.
In this case the matrices $\Pi_k$ are nothing but the projections
on $2$-dimensional eigenspaces satisfying
\be
\Pi_k^2=\Pi_k\,, \qquad
\Pi_k{}^\mu{}_\mu=2\,.
\ee
In this case we also have
\be
B_k=\frac{1}{2}E_k^{\mu\nu}F_{\mu\nu}\,.
\ee
Then we have
\be
(iF)^{2m}=\sum_{k=1}^{N}B_k^{2m}\Pi_k\,,\qquad (m\ge 1)
\ee
\be
(iF)^{2m+1}=\sum_{k=1}^{N}B_k^{2m+1}iE_k\,,\qquad
(m\ge 0)\,,
\ee
and, therefore, for
any analytic function of $tiF$ at $t=0$ we have
\be
f(tiF)=f(0)\II
+\sum_{k=1}^{N}\Bigg\{
\frac{1}{2}\bigg[f(tB_k)+f(-tB_k)-2f(0)\bigg]\Pi_k
+\frac{1}{2}\bigg[f(tB_k)-f(-tB_k)\bigg]iE_k
\Bigg\}\,.
\ee

\paragraph{Pseudo-Euclidean Case.}
This is the physically relevant case of pseudo-Eucli\-dean
(Lorentzian) metric with the signature $(-+\cdots+)$.
Then the non-zero
eigenvalues of the matrix $F$ are $\pm B_1$ (which are real)
and $\pm i B_2$, \dots, $\pm iB_N$, (which are imaginary).
We will call the invariant $B_1$,
determining the real eigenvalue, the ``electric field'' and denote it by
$B_1=E$,
and the invariants $B_k$, $k=2,\dots, N$,
 determining the imaginary eigenvalues, the
``magnetic fields''. So, in general, there is one electric field and
$(N-1)$ magnetic fields.
Again, there may be some zero eigenvalues as well.

In this case the matrices $\Pi_2$,\dots,$\Pi_N$ are the
orthogonal eigen-projections
as before, but the matrix $\Pi_1$ is equal to the negative of the
corresponding projection, in particular,
\be
\Pi_1^2=-\Pi_1\,,\qquad
\Pi_1 E_1=-E_1\,,\qquad
\Pi_1{}^\mu{}_\mu=-2\,.
\ee
Now, we have
\be
(iF)^{2m}
=-(iE)^{2m}\Pi_1
+\sum_{k=2}^{N}B_k^{2m}\Pi_k\,,\qquad (m\ge 1)
\ee
\be
(iF)^{2m+1}
=(iE)^{2m+1}E_1
+\sum_{k=2}^{N}B_k^{2m+1}iE_k\,,\qquad
(m\ge 0)\,,
\ee

Thus, to obtain the results for the pseudo-Euclidean case from the
result for the Euclidean case we should
just substitute formally
\be
B_1\mapsto iE, \qquad
iE_1\mapsto E_1,\qquad
\Pi_1\mapsto -\Pi_1\,.
\label{327xxx}
\ee
In this way, we obtain for an analytic function of $itF$,
\bea
f(tiF)&=&f(0)\II
-\frac{1}{2}\bigg[f(itE)+f(-itE)-2f(0)\bigg]\Pi_1
+\frac{1}{2}\bigg[f(itE)-f(-itE)\bigg]E_1
\nonumber\\
&&
+\sum_{k=2}^{N}\Bigg\{
\frac{1}{2}\bigg[f(tB_k)+f(-tB_k)-2f(0)\bigg]\Pi_k
+\frac{1}{2}\bigg[f(tB_k)-f(-tB_k)\bigg]iE_k
\Bigg\}\,.
\nonumber\\
\eea

\subsection{Scalar and Spinor Fields}

First of all, we note that for scalar fields
\be
\Phi^{\rm scalar}(t)=J(t)\,,\qquad
\Psi^{\rm scalar}_{\mu\nu}(t)=0\,.
\ee

For the spinor fields we have
\be
\Phi^{\rm spinor}(t)
=J(t)\;\tr\exp\left(\frac{1}{2}tiF_{\mu\nu}\gamma^{\mu\nu}\right)
\ee
\be
\Psi^{\rm spinor}_{\alpha\beta}(t)
=\frac{1}{2}\frac{\tr\;\gamma_{\alpha\beta}
\exp\left(\frac{1}{2}tiF_{\mu\nu}\gamma^{\mu\nu}\right)}
{\tr\exp\left(\frac{1}{2}tiF_{\rho\sigma}\gamma^{\rho\sigma}\right)}\,.
\ee
Here $\tr$ denotes the trace with respect to the spinor indices.

We will compute these functions as follows.
We define the matrices
\be
T_k=\frac{1}{2}iE_{k}^{\mu\nu}\gamma_{\mu\nu}\,.
\ee
Then by using the properties of the matrices $E_k$
and the product of the matrices $\gamma_{\mu\nu}$
\be
\gamma^{\mu\nu}\gamma_{\alpha\beta}=\gamma^{\mu\nu}{}_{\alpha\beta}
-4\delta^{[\mu}_{[\alpha}\gamma^{\nu]}{}_{\beta]}
-2\delta^{[\mu}_{\alpha}\delta^{\nu]}_{\beta}\II\,,
\ee
(and some other properties of Dirac matrices in $n$ dimensions)
one can show that these matrices are mutually commuting involutions,
that is,
\be
T_k^2=\II\,,
\ee
and
\be
[T_k,T_m]=0\,.
\ee
Also, the product of two different matrices is
(for $k\ne m$)
\be
T_kT_m=-\frac{1}{4}E_k^{\mu\nu}E_m^{\alpha\beta}\gamma_{\mu\nu\alpha\beta}\,.
\ee
More generally, the product of $m>1$ different matrices is
\be
T_{k_1}\cdots T_{k_m}
=\left(\frac{i}{2}\right)^m
E_{k_1}^{\mu_1\mu_2}\cdots E_{k_m}^{\mu_{2m-1}\mu_{2m}}
\gamma_{\mu_1\dots\mu_{2m}}\,.
\ee

It is well known that the matrices
$\gamma_{\mu_1\dots\mu_k}$ are traceless for any $k$ and the
trace of the product of two matrices $\gamma_{\mu_1\dots\mu_k}$
and $\gamma_{\nu_1\dots\nu_m}$ is non-zero only for $k=m$.
By using these properties we obtain the traces
\be
\tr T_k=0\,,
\ee
\be
\tr \gamma^{\alpha\beta}T_k=-2^{[n/2]}iE_k^{\alpha\beta}\,,
\ee
and for $m>1$:
\be
\tr T_{k_1}\cdots T_{k_m}=0\,,
\ee
\be
\tr \gamma^{\alpha\beta} T_{k_1}\cdots T_{k_m}=0\,.
\ee
when all indices $k_1$, \dots $k_m$ are different.

Now, by using the spectral decomposition of the matrix $F$ we easily
obtain first
\be
J(t)=\prod\limits_{k=1}^{N}\frac{tB_k}{\sinh(tB_k)}\,.
\ee
and
\be
\tr\exp\left(\frac{1}{2}tiF_{\mu\nu}\gamma^{\mu\nu}\right)
=\;\tr\prod_{k=1}^{N}
\exp\left(tT_kB_k\right)\,.
\ee
\be
\tr\gamma^{\alpha\beta}\exp\left(\frac{1}{2}tiF_{\mu\nu}\gamma^{\mu\nu}\right)
=\;\tr\gamma^{\alpha\beta}\prod_{k=1}^{N}
\exp\left(tT_kB_k\right)\,.
\ee
By using the properties of the matrices $T_k$ we get
\be
\exp\left(tT_kB_k\right)
=\cosh(tB_k)+T_k\sinh(tB_k)\,.
\ee
Therefore
\be
\tr\exp\left(\frac{1}{2}tiF_{\mu\nu}\gamma^{\mu\nu}\right)
=2^{[n/2]}
\prod\limits_{k=1}^{N}\cosh(tB_k)\,,
\ee
and
\bea
\tr \gamma^{\alpha\beta}
\prod\limits_{k=1}^{N}\exp(tT_kB_k)
&=&
\prod\limits_{j=1}^{N}\cosh(tB_j)
\sum\limits_{k=1}^{N}\tanh(tB_k)
\tr \gamma^{\alpha\beta}T_k
\nonumber\\
&=&
-2^{[n/2]}\prod\limits_{j=1}^{N}\cosh(tB_j)
\sum\limits_{k=1}^{N}\tanh(tB_k)iE_k^{\alpha\beta}\,.
\eea
Thus for the spinor fields
\be
\Phi^{\rm spinor}(t)=2^{[n/2]}\prod\limits_{k=1}^{N}
tB_k\coth(tB_k)\,,
\ee
and
\be
\Psi^{\rm spinor}_{\alpha\beta}(t)=
-\frac{1}{2}\sum\limits_{k=1}^{N}\tanh(tB_k)iE_k{}_{\alpha\beta}\,.
\ee
By the way, this simply means that
\be
\Psi^{\rm spinor}(t)=-\frac{1}{2}\tanh(tiF)\,.
\ee

\subsection{Calculation of the Tensor $V_{\mu\alpha\nu\beta}(t)$}

Next, we compute the tensor $V^{\mu\alpha}{}_{\nu\beta}(t)$.
First, we rewrite in the form
\bea
V^{\mu\alpha}{}_{\nu\beta}(t)
&=&
\left(\frac{1}{3}-\xi\right)\delta^{[\mu}_{\nu}\delta^{\alpha]}_{\beta}
+\int\limits_{0}^{1}d\tau
\Bigg\{
-\frac{1}{24}\mathcal{B}^{[\mu}{}_{[\nu}(\tau)
{\cal Z}^{\alpha]}{}_{\beta]}(\tau)
\nonumber\\
&&
+\frac{1}{16}{\cal X}^{\mu\alpha}(\tau)
\mathcal{X}_{\nu\beta}(\tau)
-\frac{1}{12}\mathcal{Y}^{[\mu}{}_{[\nu}(\tau)
\mathcal{Y}^{\alpha]}{}_{\beta]}(\tau)\Bigg\}\;,
\eea
where
\bea
{\cal X}(\tau)&=&
-\coth(tiF)+\frac{\cosh[(1-2\tau)tiF]}{\sinh(tiF)}\,,
\\[10pt]
{\cal Y}(\tau)&=&
\II+\frac{\sinh[(1-2\tau)tB_k]}{\sinh(tB_k)}\,.
\eea

Next, we parametrize these matrices as follows
\bea
\mathcal{B}(\tau)
&=&
2\tau(1-\tau)\II+
\sum_{k=1}^{N}f_{1,k}(\tau)\Pi_k\,,
\\
{\cal Z}(\tau)
&=&
4\II
+\sum_{k=1}^{N}f_{2,k}(\tau)\Pi_k\,,
\\
{\cal Y}(\tau)
&=&
2(1-\tau)\II
+\sum_{k=1}^{N}f_{3,k}(\tau)\Pi_k\,,
\\
{\cal X}(\tau)
&=&
\sum_{k=1}^{N}f_{4,k}(\tau)iE_k\,,
\\
W(t)&=&
\sum_{k=1}^{N}f_{5,k}(t)iE_k\,,
\eea
where
\bea
f_{1,k}(\tau)
&=&
\frac{\coth(tB_k)}{tB_k}
-\frac{1}{tB_k\sinh(tB_k)}\cosh[(1-2\tau)tB_k]
-2\tau(1-\tau)\;,
\\[5pt]
f_{2,k}(\tau)
&=&
3tB_k\coth(tB_k)
+\frac{tB_k}{\sinh(tB_k)}\cosh[(1-2\tau)tB_k]-4\,,
\\[5pt]
f_{3,k}(\tau)
&=&
\frac{\sinh[(1-2\tau)tB_k]}{\sinh(tB_k)}-(1-2\tau)\,,
\\[5pt]
f_{4,k}(\tau)
&=&
-\coth(tB_k)
+\frac{\cosh[(1-2\tau)tB_k]}{\sinh(tB_k)}\,,
\\[5pt]
f_{5,k}(t)
&=&
\frac{1}{2}\left(\coth(tB_k)-\frac{1}{tB_k}\right)\,.
\eea
This parametrization is convenient because all functions
$f_{m,k}(\tau)$ are analytic functions of $t$ at $t=0$ and
$f_{m,k}(\tau)\Big|_{t=0}=0$.

Then we obtain
\bea
V^{\mu\alpha}{}_{\nu\beta}(t)
&=&
\left(\frac{1}{6}-\xi\right)
\delta^{[\mu}{}{}_{[\nu}\delta^{\alpha]}{}_{\beta]}
+\sum_{k=1}^{N}
\varphi_k(t)\Pi_k{}^{[\mu}{}_{[\nu}\delta^{\alpha]}{}_{\beta]}
\nonumber\\
&&
+\sum_{k=1}^{N}\sum_{m=1}^{N}
\left[\rho_{km}(t)\Pi_k{}^{[\mu}{}_{[\nu}\Pi_{|m|}{}^{\alpha]}{}_{\beta]}
-\sigma_{km}(t)E_k{}^{\mu\alpha}E_{m\;}{}_{\nu\beta}
\right]\,,
\eea
where
\bea
\varphi_k(t)
&=&
-\frac{1}{12}\int\limits_0^1d\tau
\left[2f_{1,k}(\tau)+\tau(1-\tau)f_{2,k}(\tau)+4(1-\tau)f_{3,k}(\tau)
\right]\,,
\\
\rho_{km}(t)
&=&
-\frac{1}{48}\int\limits_0^1d\tau
\left[f_{1,k}(\tau)f_{2,m}(\tau)+f_{2,k}(\tau)f_{1,m}(\tau)
+4f_{3,k}(\tau)f_{3,m}(\tau)\right]\,,
\\
\sigma_{km}(t)
&=&
\frac{1}{16}\int\limits_0^1d\tau
f_{4,k}(\tau)f_{4,m}(\tau)\,.
\eea

\subsection{Calculation of the Coefficient Functions}

The remaining coefficient functions $\varphi_k(t)$, $\rho_{km}(t)$ and
$\sigma_{km}(t)$ are analytic functions of $t$ at $t=0$.
To compute them we use the following integrals
\bea
\int\limits_0^1 d\tau \cosh[(1-2\tau)x]
&=&
\frac{\sinh x}{x}\,,
\\
\int\limits_0^1 d\tau \sinh[(1-2\tau)x]
&=&0\,.
\eea
By differentiating these integrals with respect to $x$ we obtain all other
integrals we need
\bea
\int\limits_{0}^{1}d\tau\;\tau
\cosh[(1-2\tau)x]
&=&
\frac{1}{2}\frac{\sinh x}{x}\,,
\\
\int\limits_{0}^{1}d\tau\;\tau^2
\cosh[(1-2\tau)x]
&=&
\frac{1}{2}\left(\frac{1}{x}+\frac{1}{x^3}\right)\sinh x
-\frac{1}{2}\frac{1}{x^2}\cosh x\,,
\\
\int\limits_0^1 d\tau\;\tau \sinh[(1-2\tau)x]
&=&
-\frac{1}{2}\frac{\cosh x}{x}+\frac{1}{2}\frac{\sinh x}{x^2}\,.
\eea
We also have the integrals
\bea
\int\limits_{0}^{1}d\tau
\cosh[(1-2\tau)x]\cosh[(1-2\tau)y]
&=&
\frac{1}{2}\Bigg\{
\frac{\sinh(x+y)}{x+y}
+\frac{\sinh(x-y)}{x-y}
\Bigg\}\,,
\nonumber
\\
\\
\int\limits_{0}^{1}d\tau
\cosh[(1-2\tau)x]\sinh[(1-2\tau)y]
&=&
0\,,
\\
\int\limits_{0}^{1}d\tau
\sinh[(1-2\tau)x]\sinh[(1-2\tau)y]
&=&
\frac{1}{2}\Bigg\{
\frac{\sinh(x+y)}{x+y}
-\frac{\sinh(x-y)}{x-y}
\Bigg\}\,.
\nonumber\\
\eea

By using these integrals we obtain
\bea
\varphi_k(t)&=&
\frac{1}{6}
+\frac{3}{8}\frac{1}{(tB_k)^2}
-\frac{1}{24}\coth(tB_k)\left(tB_k+9\frac{1}{(tB_k)}\right)\,,
\\[10pt]
%
\sigma_{km}(t)&=&
\frac{1}{16}\coth(tB_k)\coth(tB_m)
-\frac{1}{16}\frac{\coth(tB_k)}{tB_m}
-\frac{1}{16}\frac{\coth(tB_m)}{tB_k}
\nonumber\\[5pt]
&&
+\frac{1}{32}\frac{\coth(tB_m)+\coth(tB_k)}{t(B_k+B_m)}
+\frac{1}{32}\frac{\coth(tB_m)-\coth(tB_k)}{t(B_k-B_m)}\,,
\\[10pt]
\rho_{km}(t)
&=&
-\frac{1}{48}
\Bigg\{
4
+9\frac{1}{(tB_k)^2}
+9\frac{1}{(tB_m)^2}
-8\frac{1}{tB_k}\coth(tB_k)
-8\frac{1}{tB_m}\coth(tB_m)
\nonumber\\[5pt]
&&
-(tB_k)\coth(tB_k)
-(tB_m)\coth(tB_m)
-3\frac{B_k}{tB_m^2}\coth(tB_k)
\nonumber\\[5pt]
&&
-3\frac{B_m}{tB_k^2}\coth(tB_m)
+3\left(\frac{B_k}{B_m}+\frac{B_m}{B_k}\right)
\coth(tB_m)\coth(tB_k)
\nonumber\\[5pt]
&&
-\frac{1}{2}\Bigg[
\frac{B_m}{B_k}
+\frac{B_k}{B_m}-4
\Bigg]
\frac{\coth(tB_m)+\coth(tB_k)}{t(B_k+B_m)}
\nonumber\\[5pt]
&&
-\frac{1}{2}\Bigg[
\frac{B_k}{B_m}
+\frac{B_m}{B_k}+4
\Bigg]\frac{\coth(tB_m)-\coth(tB_k)}{t(B_k-B_m)}
\Bigg\}\,.
\eea

\subsection{Trace of the Heat Kernel Diagonal}

The trace of the heat kernel diagonal in the general case
within the considered approximation is given by
\be
\Theta(t)\sim
(4\pi t)^{-n/2}\Phi(t)\left\{1+tB_1(t)
+\cdots\right\}\,,
\label{379xxx}
\ee
where the function $\Phi(t)$ was computed above and the coefficient
$B_1$ is given by
\bea
B_1(t)&=&
\left(\frac{1}{6}-\xi\right)R
+\sum_{k=1}^{N}
\bigg\{
\Psi^{\mu\alpha}(t)f_{5,k}(t)iE_{k}^{\nu\beta}R_{\mu\alpha\nu\beta}
+\varphi_k(t)\Pi_{k}^{\mu\nu}R_{\mu\nu}
\bigg\}
\nonumber\\
&&
+\sum_{k=1}^{N}\sum_{m=1}^{N}\bigg\{
\rho_{km}(t)\Pi_k^{\mu\nu}\Pi_m^{\alpha\beta}R_{\mu\alpha\nu\beta}
-\sigma_{km}(t)E_k^{\mu\alpha}E_m^{\nu\beta}R_{\mu\alpha\nu\beta}
\bigg\}\,.
\eea

Let us
specify it for the two cases of interest.

\subsubsection{Scalar Fields}

For scalar fields we have
\bea
\Phi^{\rm scalar}(t)&=&\prod\limits_{k=1}^{N}
\frac{tB_k}{\sinh(tB_k)}\,,
\\[10pt]
B^{\rm scalar}_1(t)&=&
\left(\frac{1}{6}-\xi\right)R
+\sum_{k=1}^{N}
\varphi_k(t)\Pi_{k}^{\mu\nu}R_{\mu\nu}
\\
&&
+\sum_{k=1}^{N}\sum_{m=1}^{N}\bigg\{
\rho_{km}(t)\Pi_k^{\mu\nu}\Pi_m^{\alpha\beta}R_{\mu\alpha\nu\beta}
-\sigma_{km}(t)E_k^{\mu\alpha}E_m^{\nu\beta}R_{\mu\alpha\nu\beta}
\bigg\}\,.
\nonumber
\eea

\subsubsection{Spinor Fields}

For the spinor fields we obtain
\bea
\Phi^{\rm spinor}(t)&=&2^{[n/2]}\prod\limits_{k=1}^{N}
tB_k\coth(tB_k)\,,
\\[10pt]
B^{\rm spinor}_1(t)&=&
-\frac{1}{12}R
+\sum_{k=1}^{N}
\varphi_k(t)\Pi_{k}^{\mu\nu}R_{\mu\nu}
\\
&&
+\sum_{k=1}^{N}\sum_{m=1}^{N}\bigg\{
\rho_{km}(t)\Pi_k^{\mu\nu}\Pi_m^{\alpha\beta}R_{\mu\alpha\nu\beta}
-\lambda_{km}(t)E_k^{\mu\alpha}E_m^{\nu\beta}R_{\mu\alpha\nu\beta}
\bigg\}\,,
\nonumber
\eea
where
\bea
\lambda_{km}(t)&=&\sigma_{km}(t)
+\frac{1}{8}\frac{\tanh(tB_m)}{tB_k}
+\frac{1}{8}\frac{\tanh(tB_k)}{tB_m}
\nonumber\\[5pt]
&&
-\frac{1}{8}\tanh(tB_m)\coth(tB_k)
-\frac{1}{8}\tanh(tB_k)\coth(tB_m)
\,.
\eea

\subsection{Equal Magnetic Fields}

We will specify the obtained result for the case when
all magnetic invariants are equal to each other, that is,
\be
B_1=\cdots=B_N=B\,.
\ee

\subsubsection{Scalar Fields}

For scalar fields it takes the form
\bea
\Phi^{\rm scalar}(t)&=&
\left(\frac{tB}{\sinh(tB)}\right)^N\,,
\\[10pt]
B^{\rm scalar}_1(t)&=&
\left(\frac{1}{6}-\xi\right)R
+\varphi(t)H^{\mu\nu}_1 R_{\mu\nu}
\nonumber
\\[5pt]
&&
+\rho(t)H^{\mu\nu}_1H^{\alpha\beta}_1R_{\mu\alpha\nu\beta}
-\sigma(t)X^{\mu\nu}_1X^{\alpha\beta}_1 R_{\mu\alpha\nu\beta}\,,
\eea
where
\be
H^{\mu\nu}_1=\sum_{k=1}^{N}\Pi_{k}^{\mu\nu}\,,
\qquad
X^{\mu\nu}_1=\sum_{k=1}^{N}E_{k}^{\mu\nu}\,,
\ee
\bea
\varphi(t)&=&
\frac{1}{6}
+\frac{3}{8}\frac{1}{(tB)^2}
-\frac{1}{24}tB\coth(tB)
-\frac{3}{8}\frac{\coth(tB)}{tB}\,,
\\[5pt]
\sigma(t)&=&
\frac{1}{16}
-\frac{3}{32}\frac{\coth(tB)}{tB}
+\frac{3}{32}\frac{1}{\sinh^2(tB)}
\\[5pt]
\rho(t)
&=&
-\frac{5}{24}
-\frac{3}{8}\frac{1}{(tB)^2}
+\frac{1}{24}tB\coth(tB)
+\frac{7}{16}\frac{\coth(tB)}{tB}
-\frac{1}{16}\frac{1}{\sinh^2(tB)}\,.
\eea

\subsubsection{Spinor Fields}

For the spinor fields we obtain
\bea
\Phi^{\rm spinor}(t)
&=&
2^{[n/2]}\left[tB\coth(tB)\right]^N\,,
\\[10pt]
B^{\rm spinor}_1(t)&=&
-\frac{1}{12}R
+\varphi(t)H^{\mu\nu}_1R_{\mu\nu}
\nonumber
\\[5pt]
&&
+\rho(t)H^{\mu\nu}_1 H^{\alpha\beta}_1R_{\mu\alpha\nu\beta}
-\lambda(t)X^{\mu\nu}_1 X^{\alpha\beta}_1
R_{\mu\alpha\nu\beta}
\,,
\eea
where
\bea
\lambda(t)&=&
-\frac{3}{16}
+\frac{3}{32}\frac{1}{\sinh^2(tB)}
+\frac{1}{4}\frac{\tanh(tB)}{tB}
-\frac{3}{32}\frac{\coth(tB)}{tB}
\,.
\eea

\subsection{Electric and Magnetic Fields}

Now we specify the above results for the
pseudo-Euclidean case when there
is one electric field and $(N-1)$ equal magnetic fields.
By using the recipe (\ref{327xxx}) we obtain the following results.

\subsubsection{Scalar Fields}

For scalar fields we have
\bea
\Phi^{\rm scalar}(t)
&=&
\frac{tE}{\sin(tE)}
\left(\frac{tB}{\sinh(tB)}\right)^{N-1}\,,
\label{3107xxx}
\\[10pt]
B^{\rm scalar}_1(t)&=&
\left(\frac{1}{6}-\xi\right)R
-\tilde\varphi(t)\Pi_{1}^{\mu\nu}R_{\mu\nu}
+\varphi(t)H_2^{\mu\nu}R_{\mu\nu}
+\tilde\rho(t)\Pi_1^{\mu\nu}\Pi_1^{\alpha\beta}R_{\mu\alpha\nu\beta}
\nonumber
\\[5pt]
&&
+\tilde\sigma(t)E_1^{\mu\alpha}E_1^{\nu\beta}R_{\mu\alpha\nu\beta}
-2\rho_1(t)H^{\mu\nu}_2\Pi_1^{\alpha\beta}
R_{\mu\alpha\nu\beta}
+2\sigma_1(t)X^{\nu\alpha}_2 E_1^{\nu\beta}R_{\mu\alpha\nu\beta}
\nonumber\\[5pt]
&&
+\rho(t)H^{\mu\nu}_2 H^{\alpha\beta}_2 R_{\mu\alpha\nu\beta}
-\sigma(t)X^{\mu\alpha}_2 X^{\nu\beta}_2 R_{\mu\alpha\nu\beta}
\,,
\eea
where
\be
H_2^{\mu\nu}=\sum_{k=2}^{N}\Pi_{k}^{\mu\nu}\,,
\qquad
X_2^{\mu\nu}=\sum_{k=2}^{N}E_{k}^{\mu\nu}\,,
\ee
\bea
\tilde\varphi(t)&=&
\frac{1}{6}
-\frac{3}{8}\frac{1}{(tE)^2}
-\frac{1}{24}tE\cot(tE)
+\frac{3}{8}\frac{\cot(tE)}{tE}\,,
\\[5pt]
\tilde\rho(t)
&=&
-\frac{5}{24}
+\frac{3}{8}\frac{1}{(tE)^2}
+\frac{1}{24}tE\cot(tE)
-\frac{7}{16}\frac{\cot(tE)}{tE}
+\frac{1}{16}\frac{1}{\sin^2(tE)}\,,
\nonumber\\
\\[5pt]
\tilde\sigma(t)&=&
\frac{1}{16}
+\frac{3}{32}\frac{\cot(tE)}{tE}
-\frac{3}{32}\frac{1}{\sin^2(tE)}\,,
\\[5pt]
\sigma_1(t)
&=&
\frac{1}{16}\cot(tE)\coth(tB)
-\frac{1}{16}\frac{\cot(tE)}{tB}
-\frac{1}{16}\frac{\coth(tB)}{tE}
\nonumber\\[10pt]
&&
+\frac{1}{16}\frac{B\cot(tE)+E\coth(tB)}{t(B^2+E^2)}
\,.
\eea
\bea
\rho_1(t)
&=&
-\frac{1}{48}
\Bigg\{
4
-9\frac{1}{(tE)^2}
+9\frac{1}{(tB)^2}
+8\frac{1}{tE}\cot(tE)
-8\frac{1}{tB}\coth(tB)
\nonumber\\[10pt]
&&
-(tE)\cot(tE)
-(tB)\coth(tB)
\nonumber\\[10pt]
&&
-3\frac{E}{tB^2}\cot(tE)
+3\frac{B}{tE^2}\coth(tB)
+3\left(\frac{E}{B}-\frac{B}{E}\right)
\coth(tB)\cot(tE)
\nonumber\\[10pt]
&&
+\frac{5B^2-E^2}{tB(B^2+E^2)}\coth(tB)
-\frac{5E^2-B^2}{tE(B^2+E^2)}\cot(tE)
\Bigg\}\,.
\eea

\subsubsection{Spinor Fields}

For the spinor fields we obtain
\bea
\Phi^{\rm spinor}(t)
&=&2^{[n/2]}tE\cot(tE)
\left[tB\coth(tB)\right]^{N-1}\,,
\label{3112xxx}
\\[10pt]
B^{\rm spinor}_1(t)&=&
-\frac{1}{12}R
-\tilde\varphi(t)\Pi_{1}^{\mu\nu}R_{\mu\nu}
+\varphi(t)H^{\mu\nu}_2R_{\mu\nu}
+\tilde\rho(t)\Pi_1^{\mu\nu}\Pi_1^{\alpha\beta}R_{\mu\alpha\nu\beta}
\nonumber
\\[5pt]
&&
+\tilde\lambda(t)E_1^{\mu\alpha}E_1^{\nu\beta}R_{\mu\alpha\nu\beta}
-2\rho_1(t)H^{\mu\nu}_2\Pi_1^{\alpha\beta}R_{\mu\alpha\nu\beta}
+2\lambda_1(t) X^{\mu\alpha}_2 E_1^{\nu\beta}R_{\mu\alpha\nu\beta}
\nonumber\\[5pt]
&&
+\rho(t)H^{\mu\nu}_2 H^{\alpha\beta}_2 R_{\mu\alpha\nu\beta}
-\lambda(t)X^{\mu\alpha}_2 X^{\nu\beta}_2 R_{\mu\alpha\nu\beta}
\,,
\eea
where
\bea
\tilde\lambda(t)&=&
-\frac{3}{16}
-\frac{3}{32}\frac{1}{\sin^2(tE)}
+\frac{1}{4}\frac{\tan(tE)}{tE}
+\frac{3}{32}\frac{\cot(tE)}{tE}\,,
\\[10pt]
\lambda_1(t)
&=&
\frac{1}{16}\cot(tE)\coth(tB)
-\frac{1}{16}\frac{\cot(tE)}{tB}
-\frac{1}{16}\frac{\coth(tB)}{tE}
\nonumber\\[10pt]
&&
+\frac{1}{16}\frac{B\cot(tE)+E\coth(tB)}{t(B^2+E^2)}
+\frac{1}{8}\frac{\tanh(tB)}{tE}
-\frac{1}{8}\frac{\tan(tE)}{tB}
\nonumber\\
&&
-\frac{1}{8}\tanh(tB)\cot(tE)
+\frac{1}{8}\tan(tE)\coth(tB)
\,.
\eea

\section{Imaginary Part of the Effective Lagrangian}
\setcounter{equation}0

Now, we can compute the imaginary part of the effective Lagrangian in the same
approximation
taking into account linear terms in the curvature.
The effective action is given by the integral over $t$ of the
trace of the heat kernel diagonal. Of course, it should be properly
regularized as discussed above. The most important point we want
to make is that in the presence of the electric field the heat kernel
is no longer a nice analytic function of $t$ but it becomes
a meromorphic function of $t$ in the complex plane of $t$ with
poles on the real axis determined by the trigonometric functions
in the coefficient functions computed above.
As was pointed out first by Schwinger these poles should be carefully
avoided by deforming the contour of integration which leads to an
imaginary part of the effective action determined by the contribution
of the residues of the poles. This imaginary part is always finite and does
not depend on the regularization.
We compute below the imaginary part of the effective Lagrangian for the scalar
and the spinor fields.

The trace of the heat kernel $\Theta(t)$ was computed above and is given
by (\ref{379xxx}).
Now, by using (\ref{232xxx})
the calculation of the
imaginary part of the effective Lagrangian is reduced to the calculation
of the residues of the functions $t^{-n/2-1}e^{-tm^2}\Phi(t)$
and $t^{-n/2}e^{-tm^2}\Phi(t)B_1(t)$ at the poles on the real line.
By using the result (\ref{3107xxx}) and (\ref{3112xxx}) for the function $\Phi$
it is not difficult to see that the function
$t^{-n/2-1}e^{-tm^2}\Phi(t)$
is a meromorphic function with
isolated simple poles at $t_{k}={k\pi}/E$
with $k=1,2\dots$. The function
$t^{-n/2}e^{-tm^2}\Phi(t)B_1(t)$ is also a meromorphic function with the same
poles but the poles could be double or even triple.
The imaginary part is, then, simply evaluated by summing the residues of the
integrand at the
poles.
It has the following form
\bea\label{gf1}
\textrm{Im}\;\mathcal{L}
&=&\pi(4\pi)^{-n/2}E^{n/2}G_{0}(x,y)
+\pi(4\pi)^{-n/2}E^{n/2-1}
\bigg[G_{1}(x,y)R\nonumber\\
&+&G_{2}(x,y)\Pi_{1}^{\mu\nu}R_{\mu\nu}
+
G_{3}(x,y)H_2{}^{\mu\nu}R_{\mu\nu}
+G_{4}(x,y)\Pi_1^{\mu\nu}\Pi_1^{\alpha\beta}R_{\mu\alpha\nu\beta}
\nonumber\\[5pt]
&+&
G_{5}(x,y)E_1^{\mu\alpha}E_1^{\nu\beta}R_{\mu\alpha\nu\beta}
+G_{6}(x,y)H_2{}^{\mu\nu}\Pi_1^{\alpha\beta}R_{\mu\alpha\nu\beta}
\nonumber\\[5pt]
&+&
G_{7}(x,y)X_2{}^{\mu\alpha} E_1^{\nu\beta}R_{\mu\alpha\nu\beta}
+G_{8}(x,y)H_2{}^{\mu\nu}H_2{}^{\alpha\beta}R_{\mu\alpha\nu\beta}
\nonumber\\
&+&
G_{9}(x,y)X_2{}^{\mu\alpha}X_2{}^{\nu\beta}
R_{\mu\alpha\nu\beta}\bigg]\;,
\eea
where
\be
x=\frac{B}{E}\;,\qquad y=\frac{m^{2}}{E}\;,
\ee
and $G_i(x,y)$ are some functions computed below.

\subsection{Scalar Fields}

At this point it is useful to introduce some auxiliary functions so that the
final result for the quantities $G^{\rm scalar}_{i}(x,y)$ can be written in a
somewhat compact form, namely
\bea
f_{k}(x,y)
&=&\left[\frac{k\pi x}{\sinh\left(k\pi
x\right)}\right]^{N-1}\exp\left({-k\pi y}\right)\;,
\\[10pt]
g_{k}(x,y)&=&(N-1)(k\pi x)\coth(k\pi x)+k\pi y\;,
\\
h_{k}(x,y)&=&\frac{1}{2}N(N-1)(k\pi x)^{2}\coth^{2}(k\pi
x)+\left(\frac{n}{2}-N\right)k\pi y
\nonumber\\
&+&\frac{1}{2}(N-1)[(n-2N)+2k\pi y](k\pi x)\coth(k\pi x)
\nonumber\\
&+&\frac{1}{2}(k\pi)^{2}[1-(N-1)x^{2}+y^{2}]\;.
\\
l_{k}(x,y)
&=&
-k\pi x+\left[\left(\frac{n}{2}-N\right)
+k\pi y\right]\coth(k\pi x)
+N(k\pi x)\coth^{2}(k\pi x)\;,
\\[10pt]
\Omega_{1,k}(x,y)
&=&\frac{1}{8}
+\frac{n-2N}{48}
-\frac{3}{8(k\pi)^{2}}\left(\frac{n}{2}-N+2\right)
\nonumber\\[5pt]
&&
+\frac{1}{24}\left(1-\frac{9}{(k\pi)^{2}}\right)g_{k}(x,y)\;,
\\[10pt]
\Omega_{2,k}(x,y)
&=&-\frac{1}{6}
-\frac{n-2N}{48}
+\frac{1}{32(k\pi)^{2}}\left(\frac{n}{2}-N+2\right)
\left(\frac{n}{2}-N+13\right)
\nonumber\\[5pt]
&-&\frac{1}{24}\left(1-\frac{21}{2(k\pi)^{2}}\right)g_{k}(x,y)
+\frac{1}{16}\frac{h_{k}(x,y)}{(k\pi)^{2}}\;,
\\[10pt]
\Omega_{3,k}(x,y)
&=&\frac{1}{16}
-\frac{3}{64(k\pi)^{2}}\left(\frac{n}{2}-N+1\right)
\left(\frac{n}{2}-N+2\right)
\nonumber\\[5pt]
&-&\frac{3}{32(k\pi)^{2}}\Big[h_{k}(x,y)+g_{k}(x,y)\Big]\;,
\\[10pt]
\Omega_{4,k}(x,y)
&=&-\frac{1}{8}
-\frac{n-2N}{48}
+\frac{3}{8(k\pi)^{2}}\left(1-\frac{1}{x^{2}}\right)
+\frac{1}{8}\left(\frac{1}{x}-x\right)\frac{l_{k}(x,y)}{k\pi}
\nonumber\\[5pt]
&+&
\frac{1}{8(k\pi)^{2}}\left[\left(\frac{n}{2}-N+1\right)
+g_{k}(x,y)\right]\left[
\frac{3x^{4}-1}{x^{2}(x^{2}+1)}\right]
\nonumber\\[5pt]
&-&\frac{1}{8}\frac{\coth(k\pi x)}{k\pi x}
\left[\frac{x^{4}-3}{x^{2}+1}-\frac{(k\pi x)^{2}}{3}\right]
-\frac{1}{24}g_{k}(x,y)\;,
\\[10pt]
\Omega_{5,k}(x,y)
&=&\frac{1}{8(k\pi)^{2}}\left[\left(\frac{n}{2}-N+1\right)
+g_{k}(x,y)\right]\frac{1}{x(x^{2}+1)}
-\frac{1}{8}\frac{l_{k}(x,y)}{(k\pi)}
\nonumber\\[5pt]
&-&
\frac{1}{8}\frac{\coth(k\pi x)}{k\pi x}
\left(\frac{x^{3}}{x^{2}+1}\right)\;.
\eea

By using these quantities we obtain the
functions $G^{\rm scalar}_{i}(x,y)$ in the form of the
following series
\bea
G^{\rm scalar}_{0}(x,y)&=&\sum_{k=1}^{\infty}
\frac{(-1)^{k+1}}{(k\pi)^{n/2}}f_{k}(x,y)\;,
\\[10pt]
G^{\rm scalar}_{1}(x,y)&=&\left(\frac{1}{6}-\xi\right)
\sum_{k=1}^{\infty}\frac{(-1)^{k+1}}{(k\pi)^{n/2-1}}f_{k}(x,y)\;,
\\[10pt]
G^{\rm scalar}_{2}(x,y)&=&-\sum_{k=1}^{\infty}
\frac{(-1)^{k+1}}{(k\pi)^{n/2-1}}f_{k}(x,y)\Omega_{1,k}(x,y)\;,
\\[10pt]
G^{\rm scalar}_{3}(x,y)&=&\sum_{k=1}^{\infty}
\frac{(-1)^{k+1}}{(k\pi)^{n/2-1}}f_{k}(x,y)\varphi(k\pi x)\;,
\\[10pt]
G^{\rm scalar}_{4}(x,y)&=&\sum_{k=1}^{\infty}
\frac{(-1)^{k+1}}{(k\pi)^{n/2-1}}f_{k}(x,y)\Omega_{2,k}(x,y)\;,
\\[10pt]
G^{\rm scalar}_{5}(x,y)&=&\sum_{k=1}^{\infty}
\frac{(-1)^{k+1}}{(k\pi)^{n/2-1}}f_{k}(x,y)\Omega_{3,k}(x,y)\;,
\\[10pt]
G^{\rm scalar}_{6}(x,y)&=&-\sum_{k=1}^{\infty}
\frac{(-1)^{k+1}}{(k\pi)^{n/2-1}}f_{k}(x,y)\Omega_{4,k}(x,y)\;,
\\[10pt]
G^{\rm scalar}_{7}(x,y)&=&\sum_{k=1}^{\infty}
\frac{(-1)^{k+1}}{(k\pi)^{n/2-1}}f_{k}(x,y)\Omega_{5,k}(x,y)\;,
\\[10pt]
G^{\rm scalar}_{8}(x,y)&=&\sum_{k=1}^{\infty}
\frac{(-1)^{k+1}}{(k\pi)^{n/2-1}}f_{k}(x,y)\rho(k\pi x)\;,
\\[10pt]
G^{\rm scalar}_{9}(x,y)&=&\sum_{k=1}^{\infty}
\frac{(-1)^{k+1}}{(k\pi)^{n/2-1}}f_{k}(x,y)\sigma(k\pi x)\;.
\eea

\subsection{Spinor Fields}

Exactly as we did in the previous section, we introduce, now, some auxiliary
functions that will be useful in the presentation of the final result,
namely
\bea
f_{S,k}(x,y)
&=&
\Big[(k\pi x)\coth(k\pi x)\Big]^{N-1}\exp\left({-k\pi y}\right)\;,
\\[10pt]
g_{S,k}(x,y)&=&
(N-1)(k\pi x)\coth(k\pi x)-(N-1)(k\pi x)\tanh(k\pi x)+k\pi y\,,
\\[10pt]
h_{S,k}(x,y)&=&\frac{1}{2}(k\pi y)^{2}
-(N-1)^{2}(k\pi x)^{2}
+\left(\frac{n}{2}-N\right)k\pi y
\\[5pt]
&+&
\frac{1}{2}(N-1)\left(n-2N
+2k\pi y\right)(k\pi x)\Big[\coth(k\pi x)
-\tanh(k\pi x)\Big]
\nonumber\\[5pt]
&+&
\frac{1}{2}N(N-1)(k\pi x)^{2}\coth^{2}(k\pi x)
+\frac{1}{2}(N-1)(N-2)(k\pi x)^{2}\tanh^{2}(k\pi x)\,,
\nonumber
\\[10pt]
l_{S,k}(x,y)
&=&
-Nk\pi x+\left(\frac{n}{2}-N+k\pi y\right)
\coth(k\pi x)+N(k\pi x)\coth^{2}(k\pi x)\;,
\\[10pt]
p_{S,k}(x,y)&=&
(N-2)k\pi x+\left(\frac{n}{2}-N+k\pi y\right)
\tanh(k\pi x)-(N-2)(k\pi x)\tanh^{2}(k\pi x)\;,
\nonumber\\
&&
\\
\Lambda_{1,k}(x,y)&=&\frac{1}{8}
+\frac{n-2N}{48}
-\frac{3}{8(k\pi)^{2}}\left(\frac{n}{2}-N+2\right)
+\frac{1}{24}\left(1
-\frac{9}{(k\pi)^{2}}\right)g_{S,k}(x,y)\;,
\nonumber\\
\\
\Lambda_{2,k}(x,y)&=&-\frac{1}{6}
-\frac{n-2N}{48}
+\frac{1}{32(k\pi)^{2}}\left(\frac{n}{2}-N+2\right)
\left(\frac{n}{2}-N+13\right)
\nonumber\\[5pt]
&-&\frac{1}{24}\left(1-\frac{21}{2(k\pi)^{2}}\right)g_{S,k}(x,y)
+\frac{1}{16}\frac{h_{S,k}(x,y)}{(k\pi)^{2}}\;,
\\[10pt]
\Lambda_{3,k}(x,y)&=&-\frac{3}{16}
-\frac{3}{64(k\pi)^{2}}\left(\frac{n}{2}-N+1\right)
\left(\frac{n}{2}-N+2\right)
\nonumber\\[5pt]
&-&\frac{3}{32(k\pi)^{2}}\Big[g_{S,k}(x,y)+h_{S,k}(x,y)\Big]\;,
\eea
\bea
\Lambda_{4,k}(x,y)&=&-\frac{1}{8}
-\frac{n-2N}{48}
+\frac{3}{8(k\pi)^{2}}\left(1-\frac{1}{x^{2}}\right)
+\frac{1}{8}\left(\frac{1}{x}-x\right)\frac{l_{S,k}(x,y)}{k\pi}
\nonumber\\[5pt]
&+&\frac{1}{8(k\pi)^{2}}\left(\frac{n}{2}-N+1
+g_{S,k}(x,y)\right)
\frac{3x^{4}-1}{x^{2}(x^{2}+1)}
\nonumber\\[5pt]
&-&\frac{1}{8}\frac{\coth(k\pi x)}{k\pi x}
\left[\frac{x^{4}-3}{x^{2}+1}-\frac{(k\pi x)^{2}}{3}\right]
-\frac{1}{24}g_{S,k}(x,y)\;,
\\[10pt]
\Lambda_{5,k}(x,y)
&=&\frac{1}{8(k\pi)^{2}}\left(\frac{n}{2}-N+1
+g_{S,k}(x,y)\right)
\frac{1}{x(x^{2}+1)}
-\frac{1}{8}\frac{l_{S,k}(x,y)}{k\pi}
\nonumber\\
&-&\frac{1}{8}\frac{\coth(k\pi x)}{k\pi x}
\left(\frac{x^{3}}{x^{2}+1}\right)
+\frac{1}{4(k\pi)}\Big[\tanh(k\pi x)
+p_{S,k}(x,y)\Big]\;.
\eea

By using the above functions we can write the explicit expression for the
quantities
$G^{\rm spinor}_{i}(x,y)$
\bea
G^{\rm spinor}_{0}(x,y)&=&2^{\left[\frac{n}{2}\right]}
\sum_{k=1}^{\infty}
\frac{1}{(k\pi)^{n/2}}f_{S,k}(x,y)\;,
\\[10pt]
G^{\rm spinor}_{1}(x,y)&=&-\frac{2^{\left[\frac{n}{2}\right]}}{12}
\sum_{k=1}^{\infty}
\frac{1}{(k\pi)^{n/2-1}}f_{S,k}(x,y)\;,
\\[10pt]
G^{\rm spinor}_{2}(x,y)&=&-2^{\left[\frac{n}{2}\right]}
\sum_{k=1}^{\infty}
\frac{1}{(k\pi)^{n/2-1}}f_{S,k}(x,y)\Lambda_{1,k}(x,y)\;,
\\[10pt]
G^{\rm spinor}_{3}(x,y)&=&2^{\left[\frac{n}{2}\right]}
\sum_{k=1}^{\infty}
\frac{1}{(k\pi)^{n/2-1}}f_{S,k}(x,y)\varphi(k\pi x)\;,
\\[10pt]
G^{\rm spinor}_{4}(x,y)&=&2^{\left[\frac{n}{2}\right]}
\sum_{k=1}^{\infty}
\frac{1}{(k\pi)^{n/2-1}}f_{S,k}(x,y)\Lambda_{2,k}(x,y)\;,
\\[10pt]
G^{\rm spinor}_{5}(x,y)&=&2^{\left[\frac{n}{2}\right]}
\sum_{k=1}^{\infty}
\frac{1}{(k\pi)^{n/2-1}}f_{S,k}(x,y)\Lambda_{3,k}(x,y)\;,
\\[10pt]
G^{\rm spinor}_{6}(x,y)&=&-2^{\left[\frac{n}{2}\right]}
\sum_{k=1}^{\infty}
\frac{1}{(k\pi)^{n/2-1}}f_{S,k}(x,y)\Lambda_{4,k}(x,y)\;,
\eea
\bea
G^{\rm spinor}_{7}(x,y)&=&2^{\left[\frac{n}{2}\right]}
\sum_{k=1}^{\infty}
\frac{1}{(k\pi)^{n/2-1}}f_{S,k}(x,y)\Lambda_{5,k}(x,y)\;,
\\[10pt]
G^{\rm spinor}_{8}(x,y)&=&2^{\left[\frac{n}{2}\right]}
\sum_{k=1}^{\infty}
\frac{1}{(k\pi)^{n/2-1}}f_{S,k}(x,y)\rho(k\pi x)\;,
\\[10pt]
G^{\rm spinor}_{9}(x,y)&=&-2^{\left[\frac{n}{2}\right]}
\sum_{k=1}^{\infty}
\frac{1}{(k\pi)^{n/2-1}}f_{S,k}(x,y)\lambda(k\pi x)\;.
\eea

Notice that because of the infrared cutoff factor $e^{-k\pi y}$ the functions
$G_i(x,y)$ are exponentially small for massive fields in weak electric fields
when the parameter is large, $y>>1$ (that is, $m^2>>E$), independently on $x$.
In this case, all
these functions are
approximated by just the first term of the series corresponding to
$k=1$.


\section{Strong Electric Field in Four Dimensions}

The formulas obtained in the previous section are very general and are valid in
any dimensions.
In this section we will present some particular cases of major interest.

\subsection{Four Dimensions}

In this section we will consider the physical case when $n=4$. Obviously in
four dimensions
we only have two invariants, and, therefore, $N=2$.
The imaginary part of the effective Lagrangian reads now
\bea\label{gf1}
\textrm{Im}\;\mathcal{L}
&=&\pi(4\pi)^{-2}E^{2}G_{0}(x,y)
+\pi(4\pi)^{-2}E
\bigg[G_{1}(x,y)R\nonumber\\
&+&G_{2}(x,y)\Pi_{1}^{\mu\nu}R_{\mu\nu}
+
G_{3}(x,y)\Pi_2{}^{\mu\nu}R_{\mu\nu}
+G_{4}(x,y)\Pi_1^{\mu\nu}\Pi_1^{\alpha\beta}R_{\mu\alpha\nu\beta}
\nonumber\\[5pt]
&+&
G_{5}(x,y)E_1^{\mu\alpha}E_1^{\nu\beta}R_{\mu\alpha\nu\beta}
+G_{6}(x,y)\Pi_2{}^{\mu\nu}\Pi_1^{\alpha\beta}R_{\mu\alpha\nu\beta}
\nonumber\\[5pt]
&+&
G_{7}(x,y)E_2{}^{\mu\alpha} E_1^{\nu\beta}R_{\mu\alpha\nu\beta}
+G_{8}(x,y)\Pi_2{}^{\mu\nu}\Pi_2{}^{\alpha\beta}R_{\mu\alpha\nu\beta}
\nonumber\\
&+&
G_{9}(x,y)E_2{}^{\mu\alpha}E_2{}^{\nu\beta}
R_{\mu\alpha\nu\beta}\bigg]\;.
\eea

For scalar fields in four dimensions the functions $G^{\rm scalar}_{i}(x,y)$
take the form
\bea
G^{\rm scalar}_{0}(x,y)
&=&\frac{x}{\pi}
\sum_{k=1}^{\infty}
\frac{e^{-k\pi y}}{k\,\sinh(k\pi x)}\;,
\\[10pt]
G^{\rm scalar}_{1}(x,y)
&=&\left(\frac{1}{6}-\xi\right)x\sum_{k=1}^{\infty}
\frac{e^{-k\pi y}}{\sinh(k\pi x)}\;,
\\[10pt]
G^{\rm scalar}_{2}(x,y)
&=&-x\sum_{k=1}^{\infty}
\frac{e^{-k\pi y}}{\sinh(k\pi x)}
\Bigg\{\frac{1}{8}
-\frac{3}{4(k\pi)^{2}}
+\frac{1}{24}\left(k\pi-\frac{9}{k\pi}\right)
\left[y+x \coth(k\pi x)\right]\Bigg\}\;,
\nonumber\\
\\
G^{\rm scalar}_{3}(x,y)
&=&
x\sum_{k=1}^{\infty}
\frac{e^{-k\pi y}}{\sinh(k\pi x)}
\Bigg\{\frac{1}{6}
+\frac{3}{8}\frac{1}{(k\pi x)^2}
-\frac{1}{24}k\pi x\coth(k\pi x)
-\frac{3}{8}\frac{\coth(k\pi x)}{k\pi x}
\Bigg\}
\;,
\nonumber\\
\\
G^{\rm scalar}_{4}(x,y)
&=&x\sum_{k=1}^{\infty}
\frac{e^{-k\pi y}}{\sinh(k\pi x)}
\Bigg\{-\frac{13}{96}
+\frac{13}{16(k\pi)^{2}}
-\frac{x^{2}}{32}
+\frac{y^{2}}{32}
+\left(\frac{7}{16(k\pi)}-\frac{k\pi}{24}\right)y
\nonumber\\[5pt]
&&
+\left(\frac{y}{16}+\frac{7}{16 k\pi}-\frac{k\pi}{24}\right)
x\coth(k\pi x)
+\frac{x^2}{16}\coth^2(k\pi x)\Bigg\}\,,
\\[10pt]
G^{\rm scalar}_{5}(x,y)&=&
x\sum_{k=1}^{\infty}
\frac{e^{-k\pi y}}{\sinh(k\pi x)}
\Bigg\{\frac{1}{64}
-\frac{3}{32(k\pi)^{2}}
+\frac{3x^{2}}{64}
-\frac{3y^{2}}{64}
-\frac{3}{32k\pi}y
\nonumber\\[5pt]
&-&\frac{3}{32}
\left(y+\frac{1}{k\pi}\right)\coth(k\pi x)
-\frac{3}{32}x^2\coth^2(k\pi x)\Bigg\}\,,
\\[10pt]
G^{\rm scalar}_{6}(x,y)&=&
-\frac{x}{x^2+1}\sum_{k=1}^{\infty}
\frac{e^{-k\pi y}}{\sinh(k\pi x)}\Bigg\{-\frac{1}{4}
-\frac{1}{2(k\pi x)^{2}}
+\frac{3x^{2}}{4(k\pi)^{2}}
+\frac{x^{2}}{8}(x^{2}-1)
\nonumber\\[5pt]
&-&\frac{k\pi}{24}y(x^2+1)
-\frac{1}{8k\pi}y\left(\frac{1}{x^{2}}-3x^{2}\right)
-\frac{1}{4}(x^{4}-1)\coth^{2}(k\pi x)
\nonumber\\[5pt]
&+&\left[\frac{1}{4k\pi}\left(x^{3}+\frac{1}{x}\right)
+\frac{y}{8}\left(\frac{1}{x}-x^{3}\right)\right]\coth(k\pi x)
\Bigg\}\;,
\eea
\bea
G^{\rm scalar}_{7}(x,y)&=&
\frac{1}{x^2+1}
\sum_{k=1}^{\infty}
\frac{e^{-k\pi y}}{\sinh(k\pi x)}
\Bigg\{
\frac{1}{8(k\pi)^{2}}
+\frac{x^{2}}{8}(x^{2}+1)
+\frac{y}{8k\pi}
\nonumber\\[5pt]
&-&\frac{x}{8}\left[y(1+x^2)
-\frac{1}{k\pi}(1-x^2)\right]\coth(k\pi x)
\nonumber\\[5pt]
&-&\frac{1}{4}x^{2}(x^{2}+1)\coth^{2}(k\pi x)\Bigg\}\;,
\\[10pt]
G^{\rm scalar}_{8}(x,y)
&=&
x\sum_{k=1}^{\infty}
\frac{e^{-k\pi y}}{\sinh(k\pi x)}
\Bigg\{
-\frac{7}{48}
-\frac{3}{8}\frac{1}{(k\pi x)^2}
-\frac{1}{16}\coth^2(k\pi x)
\nonumber\\[5pt]
&&
+\left(\frac{k\pi}{24}x
+\frac{7}{16k\pi x}\right)\coth(k\pi x)
\Bigg\}
\;,
\\[10pt]
G^{\rm scalar}_{9}(x,y)
&=&x\sum_{k=1}^{\infty}
\frac{e^{-k\pi y}}{\sinh(k\pi x)}
\Bigg\{
-\frac{1}{32}
-\frac{3}{32 k\pi x}\coth(k\pi x)
+\frac{3}{32}\coth^2(k\pi x)
\Bigg\}
\;.
\nonumber\\
\eea

For spinor fields in four dimensions the functions
$G^{\rm spinor}_{i}(x,y)$ take the form
\bea
G^{\rm spinor}_{0}(x,y)
&=&\frac{4x}{\pi}
\sum_{k=1}^{\infty}
\frac{1}{k}\coth(k\pi x)e^{-k\pi y}\;,
\\[10pt]
G^{\rm spinor}_{1}(x,y)
&=&-\frac{x}{3}
\sum_{k=1}^{\infty}
\coth(k\pi x)e^{-k\pi y}\;,
\\[10pt]
G^{\rm spinor}_{2}(x,y)
&=&-4x
\sum_{k=1}^{\infty}
\coth(k\pi x)e^{-k\pi y}\Bigg\{\frac{1}{8}
-\frac{3}{4(k\pi)^{2}}
\nonumber\\[5pt]
&+&\frac{1}{24}\left(k\pi
-\frac{9}{k\pi}\right)\Big[y+x\coth(k\pi x)-x
\tanh(k\pi x)\Big]\Bigg\}\;,
\\[10pt]
G^{\rm spinor}_{3}(x,y)
&=&4
x\sum_{k=1}^{\infty}
\coth(k\pi x)e^{-k\pi y}
\Bigg\{\frac{1}{6}
+\frac{3}{8}\frac{1}{(k\pi x)^2}
-\frac{1}{24}k\pi x\coth(k\pi x)
\nonumber\\[5pt]
&&
-\frac{3}{8}\frac{\coth(k\pi x)}{k\pi x}
\Bigg\}
\;,
\eea
\bea
G^{\rm spinor}_{4}(x,y)&=&
4x\sum_{k=1}^{\infty}
\coth(k\pi x)e^{-k\pi y}\Bigg\{-\frac{1}{6}
+\frac{13}{16(k\pi)^{2}}
-\frac{x^{2}}{16}
+\frac{y^{2}}{32}
\nonumber\\[5pt]
&-&\frac{y}{24}\left(k\pi-\frac{21}{2k\pi}\right)
+\frac{x^{2}}{16}\coth^{2}(k\pi x)
\nonumber\\[5pt]
&-&\frac{x}{24}\left(k\pi
-\frac{21}{2k\pi}-\frac{3y}{2}\right)\Big[\coth(k\pi x)
-\tanh(k\pi x)\Big]\Bigg\}\;,
\\[10pt]
%
G^{\rm spinor}_{5}(x,y)
&=&4x
\sum_{k=1}^{\infty}
\coth(k\pi x)e^{-k\pi y}
\Bigg\{-\frac{3}{16}
-\frac{3}{32(k\pi)^{2}}
+\frac{3x^{2}}{32}
\nonumber\\[5pt]
&-&\frac{3y^{2}}{64}
-\frac{3y}{32(k\pi)}
-\frac{3x^{2}}{32}\coth^{2}(k\pi x)
\nonumber\\[5pt]
&-&\frac{3x}{32}\left(\frac{1}{k\pi}+y\right)
\Big[\coth(k\pi x)-\tanh(k\pi x)\Big]\Bigg\}\;,
\\[10pt]
G^{\rm spinor}_{6}(x,y)
&=&-\frac{4x}{x^2+1}
\sum_{k=1}^{\infty}
\coth(k\pi x)e^{-k\pi y}\Bigg\{-\frac{3}{8}
+\frac{3}{4(k\pi)^{2}}\left(x^{2}-\frac{2}{3x^{2}}\right)
\nonumber\\[5pt]
&+&\frac{x^{2}}{8}(2x^{2}-1)
-\frac{x^{2}y}{24}\left(k\pi-\frac{9}{k\pi}\right)
-\frac{y}{24}\left(k\pi+\frac{3}{k\pi x^{2}}\right)
\nonumber\\[5pt]
&+&\left[\frac{1}{4k\pi x}+\frac{y}{8x}-\frac{x^{3}}{8}
\left(y-\frac{2}{k\pi}
\right)\right]\coth(k\pi x)
\nonumber\\[5pt]
&+&\left[\frac{1}{8k\pi x}+\frac{k\pi x}{24}
+\frac{x^{3}}{24}\left(k\pi-\frac{9}{k\pi}\right)
\right]\tanh(k\pi x)
\nonumber\\[5pt]
&-&\frac{1}{4}(x^{4}-1)\coth^{2}(k\pi x)
\Bigg\}\;,
\\[10pt]
G^{\rm spinor}_{7}(x,y)
&=&\frac{4}{x^2+1}
\sum_{k=1}^{\infty}
\coth(k\pi x)
e^{-k\pi y}\Bigg\{\frac{1}{8(k\pi)^{2}}
+\frac{x^{2}}{4}(x^{2}+1)
\nonumber\\[5pt]
&+&\frac{y}{8k\pi}
-\frac{x^{2}}{4}(x^{2}+1)\coth^{2}(k\pi x)
\nonumber\\
&+&\frac{x}{8}\left[\frac{1}{k\pi}(1-x^{2})-y(1+x^{2})
\right]\coth(k\pi x)
\nonumber\\[5pt]
&+&\frac{x}{8}\left[\frac{1}{k\pi}(1+2x^{2})+2y(1+x^{2})
\right]\tanh(k\pi x)\Bigg\}\;,
\eea
\bea
G^{\rm spinor}_{8}(x,y)
&=&4x
\sum_{k=1}^{\infty}
\coth(k\pi x)e^{-k\pi y}
\Bigg\{
-\frac{7}{48}
-\frac{3}{8}\frac{1}{(k\pi x)^2}
+\frac{1}{24}k\pi x\coth(k\pi x)
\nonumber\\[5pt]
&&
+\frac{7}{16}\frac{\coth(k\pi x)}{k\pi x}
-\frac{1}{16}\coth^2(k\pi x)
\Bigg\}
\;,
\\[10pt]
G^{\rm spinor}_{9}(x,y)
&=&-4x
\sum_{k=1}^{\infty}
\coth(k\pi x)e^{-k\pi y}
\Bigg\{
-\frac{9}{32}
+\frac{3}{32}\coth^2(k\pi x)
+\frac{1}{4}\frac{\tanh(k\pi x)}{k\pi x}
\nonumber\\[5pt]
&&
-\frac{3}{32}\frac{\coth(k\pi x)}{k\pi x}
\Bigg\}
\;,
\eea
%

\subsection{Supercritical Electric Field}

As we already mentioned above, the functions $G_i(x,y)$ are exponentially
small for massive fields in weak electric fields for large $y=m^2/E$, as $y\to
\infty$. Now we are considering the opposite case of light (or massless)
fields in strong (supercritical) electric fields, when $y\to 0$ with a fixed
$x$. This corresponds to the regime
\be
m^2<<B, E\,.
\ee

\subsubsection{Scalar Fields}

The infrared (massless) limit for scalar fields is regular---there
are no infrared divergences. This is due
to the presence of the hyperbolic sine $\sinh(k\pi x)$ in the denominator, which gives a cut-off for large $k$ in the series, and
therefore, assures its convergence. The result for the massless limit in the
scalar case can be simply obtained by setting $y=0$ in the above formulas
for the functions $G_i(x,y)$.

\subsubsection{Spinor Fields}

The spinor case is quite different. The presence of the hyperbolic cotangent
$\coth(k\pi x)$ does not provide a cut-off for the
convergence of the series as $k\to\infty$. This leads, in the spinor case in
four dimensions, to the presence of infrared divergences as $y=m^2/E\to 0$.
By carefully studying the behavior of the series as $k\to \infty$ for a finite
$y$ and then letting $y\to 0$ we compute the asymptotic expansion of the
functions $G_{i}^{\rm spinor}(x,y)$ as $y\to 0$.

We obtain
\bea
G_{0}^{\rm spinor}(x,y)
&=&\frac{2x}{3}+O(y)\;,
\label{564xxx}
\\[5pt]
G_{1}^{\rm spinor}(x,y)
&=&-\frac{1}{3\pi}\frac{x}{y}
+\frac{x}{8}+O(y)\;,
\\[5pt]
G_{2}^{\rm spinor}(x,y)
&=&-\frac{2}{3\pi}\frac{x}{y}
+\frac{3x}{4}+O(y)\;,
\\[5pt]
G_{3}^{\rm spinor}(x,y)
&=&-\frac{1}{6\pi}\frac{x^{2}}{y^{2}}
+\frac{2}{3\pi}\frac{x}{y}
+\frac{3}{2\pi}\log(\pi y)
+\frac{x^{2}(\pi x-24)+18}{72x}+O(y)\;,
\nonumber\\
\\
G_{4}^{\rm spinor}(x,y)
&=&-\frac{5}{6\pi}\frac{x}{y}
+\frac{7x}{8}+O(y)\;,
\\[5pt]
G_{5}^{\rm spinor}(x,y)
&=&-\frac{3}{4\pi}\frac{x}{y}
+\frac{5x}{16}+O(y)\;,
\\[5pt]
G_{6}^{\rm spinor}(x,y)
&=&\frac{x^{2}-1}{6\pi(x^{2}+1)}\frac{x^{2}}{y^{2}}
+\frac{2}{3\pi}\frac{x}{y}
-\frac{x^{4}-3}{2\pi(x^{2}+1)}\log(\pi y)
\\[5pt]
&-&\frac{6\pi(9x^{4}+3x^{2}-4)
-36x(x^{4}-1)+\pi^{2}x^{3}(x^{2}-1)}{72\pi
x(x^{2}+1)}+O(y)\;,
\nonumber
\\[5pt]
G_{7}^{\rm spinor}(x,y)
&=&-\frac{x(x^{2}+2)}{2\pi(x^{2}+1)}\log(\pi y)
+\frac{6x(x^{2}+1)+\pi}{12\pi(x^{2}+1)}+O(y)\;,
\\[5pt]
G_{8}^{\rm spinor}(x,y)
&=&\frac{1}{6\pi}\frac{x^{2}}{y^{2}}
-\frac{5}{6\pi}\frac{x}{y}
-\frac{7}{4\pi}\log(\pi y)
-\frac{x^{2}(\pi x-30)+18}{72x}+O(y)\;,
\nonumber\\
\\[5pt]
G_{9}^{\rm spinor}(x,y)
&=&\frac{3}{4\pi}\frac{x}{y}
+\frac{5}{8\pi}\log(\pi y)
-\frac{3x}{8}+O(y)\;.
\label{573xxx}
\eea

Thus, we clearly see the infrared divergences of order $x^2/y^2=B^2/m^{4}$,
$x/y=B/m^{2}$ and $\log y=\log(m^2/E)$.

\subsection{Pure Electric Field}

We analyze now the case of pure electric field without a
magnetic field, that is, $B=0$, which corresponds to the limit $x\to 0$
with fixed $y$. This corresponds to the physical regime when
\be
B<<m^2, E\,.
\ee
In this discussion we present the results
in arbitrary
dimension first and then we specialize them to the physical dimension $n=4$.

\subsubsection{Scalar Fields}

We now evaluate the functions $G_i(x,y)$ for $x=0$ and a finite $y$.
In this limit we are presented with
series of the following general form
\be
\chi_{n}^{\rm scalar}(y)=\sum_{k=1}^{\infty}
\frac{(-1)^{k+1}e^{-k\pi y}}{k^{n/2}}\;.
\ee
This series can be expressed in terms of the
polylogarithmic
function defined by
\be
\textrm{Li}_{j}(z)=\sum_{k=1}^{\infty}\frac{z^{k}}{k^{j}}\;,
\ee
so that, we have
\be\label{7}
\chi_{n}^{\rm scalar}(y)
=-\textrm{Li}_{\frac{n}{2}}(-e^{-\pi y})\;.
\ee

It is not difficult to notice that the limit as $x\rightarrow 0$ of the
functions
$G_{3}^{\rm scalar}$, $G_{6}^{\rm scalar}$, $G_{7}^{\rm scalar}$,
$G_{8}^{\rm scalar}$ and $G_{9}^{\rm scalar}$ vanish
identically, that is,
\be
G_{3}^{\rm scalar}(0,y)=G_{6}^{\rm scalar}(0,y)=
G_{7}^{\rm scalar}(0,y)=G_{8}^{\rm scalar}(0,y)=
G_{9}^{\rm scalar}(0,y)=0\,.
\ee
The explicit expression for the remaining non-vanishing $G_{i}^{\rm scalar}$
for pure
electric field in $n$ dimensions is
\bea
G_{0}^{\rm scalar}(0,y)&=&
-\pi^{-n/2}
\textrm{Li}_{\frac{n}{2}}(-e^{-\pi y})\;,
\\[10pt]
G_{1}^{\rm scalar}(0,y)&=&
-\left(\frac{1}{6}-\xi\right)\frac{1}{\pi^{n/2-1}}
\textrm{Li}_{\frac{n}{2}-1}(-e^{-\pi y})\;,
\\
G_{2}^{\rm scalar}(0,y)&=&
-\frac{1}{48\pi^{n/2+1}}
\bigg\{2\pi^{3}y\textrm{Li}_{\frac{n}{2}-2}(-e^{-\pi y})
+(n+4)\pi^{2}\textrm{Li}_{\frac{n}{2}-1}(-e^{-\pi y})
\nonumber\\[10pt]
&-&18\pi y\textrm{Li}_{\frac{n}{2}}(-e^{-\pi y})
-9(n+2)\textrm{Li}_{\frac{n}{2}+1}(-e^{-\pi y})\bigg\}\;,
\eea
\bea
G_{4}^{\rm scalar}(0,y)&=&
\frac{1}{384\pi^{n/2+1}}
\bigg\{-16\pi^{3}y\textrm{Li}_{\frac{n}{2}-2}(-e^{-\pi y})
-4\pi^{2}(2n+9-3y^{2})\textrm{Li}_{\frac{n}{2}-1}(-e^{-\pi y})
\nonumber\\[5pt]
&+&12(n+12)\pi y\textrm{Li}_{\frac{n}{2}}(-e^{-\pi y})
+3(n+2)(n+24)\textrm{Li}_{\frac{n}{2}+1}(-e^{-\pi y})\bigg\}\;,
\\[10pt]
G_{5}^{\rm scalar}(0,y)&=&\frac{1}{256\pi^{n/2+1}}
\bigg\{4\pi^{2}(1-3y^{2})\textrm{Li}_{\frac{n}{2}-1}(-e^{-\pi y})
-12n\pi y\textrm{Li}_{\frac{n}{2}}(-e^{-\pi y})
\nonumber\\[5pt]
&-&3n(n+2)\textrm{Li}_{\frac{n}{2}+1}(-e^{-\pi y})\bigg\}\;.
\eea

In the physical case of $n=4$ some of the polylogarithmic functions
can be expressed in terms of elementary functions. In this case we have
\bea
G_{0}^{\rm scalar}(0,y)
&=&-\frac{1}{\pi^{2}}\textrm{Li}_{2}(-e^{-\pi
y})\;.
\\[10pt]
G_{1}^{\rm scalar}(0,y)&=&
-\left(\frac{1}{6}-\xi\right)
\frac{1}{\pi}\ln(1+e^{-\pi y})\;,
\\[10pt]
G_{2}^{\rm scalar}(0,y)&=&
\frac{1}{48\pi^{3}}
\bigg\{\frac{2\pi^{3}y e^{-\pi y}}{1+e^{-\pi y}}
+8\pi^{2}\ln(1+e^{-\pi y})
\nonumber\\[5pt]
&+&18\pi y\textrm{Li}_{2}(-e^{-\pi y})
+54\textrm{Li}_{3}(-e^{-\pi y})\bigg\}\;,
\\[10pt]
G_{4}^{\rm scalar}(0,y)&=&
\frac{1}{384\pi^{3}}
\bigg\{\frac{16\pi^{3}y e^{-\pi y}}{1+e^{-\pi y}}
+4\pi^{2}(17-3y^{2})\ln(1+e^{-\pi y})
\nonumber\\[5pt]
&+&192\pi y\textrm{Li}_{2}(-e^{-\pi y})
+504\textrm{Li}_{3}(-e^{-\pi y})\bigg\}\;,
\\[10pt]
G_{5}^{\rm scalar}(0,y)&=&
-\frac{1}{256\pi^{3}}
\bigg\{4\pi^{2}(1-3y^{2})\ln(1+e^{-\pi y})
+48\pi y\textrm{Li}_{2}(-e^{-\pi y})
\nonumber\\[5pt]
&+&72\textrm{Li}_{3}(-e^{-\pi y})\bigg\}\;.
\eea


We study now the behavior of these functions as $y\to 0$, which corresponds to
the limit
\be
B=0\,,\qquad m^2<<E\,.
\ee
By taking the
limit as $y\rightarrow 0$ of the expression (\ref{7}) and by noticing that
\be\label{38}
\textrm{Li}_{n}(-1)=-(1-2^{1-n})\zeta\left(n\right)\;,
\ee
where $\zeta(x)$ denotes the Riemann zeta function, we obtain
\be\label{76}
G_{0}^{\rm scalar}(0,0)
=\frac{(1-2^{1-n/2})}{\pi^{n/2}}
\zeta\left(\frac{n}{2}\right)\;.
\ee
Next, by taking the limit
as $y\rightarrow 0$
and by using the formula
(\ref{38}),
it is not difficult to obtain
\bea
G_{1}^{\rm scalar}(0,0)&=&
-\left(\frac{1}{6}-\xi\right)\pi^{1-n/2}(1-2^{2-n/2})
\zeta\left(\frac{n}{2}-1\right)\;,
\\[10pt]
G_{2}^{\rm scalar}(0,0)
&=&-\frac{1}{48\pi^{n/2+1}}
\bigg\{-(n+4)\pi^{2}(1-2^{2-n/2})\zeta\left(\frac{n}{2}-1\right)
\nonumber\\[5pt]
&+&9(n+2)(1-2^{-n/2})\zeta\left(\frac{n}{2}+1\right)\bigg\}\;,
\\[10pt]
G_{4}^{\rm scalar}(0,0)
&=&\frac{1}{384\pi^{n/2+1}}
\bigg\{4\pi^{2}(2n+9)(1-2^{2-n/2})\zeta\left(\frac{n}{2}-1\right)
\nonumber\\[5pt]
&-&3(n+2)(n+24)(1-2^{-n/2})\zeta\left(\frac{n}{2}+1\right)\bigg\}\;,
\\[10pt]
G_{5}^{\rm scalar}(0,0)
&=&\frac{1}{256\pi^{n/2+1}}
\bigg\{-4\pi^{2}(1-2^{2-n/2})\zeta\left(\frac{n}{2}-1\right)
\nonumber\\[5pt]
&+&3n(n+2)(1-2^{-n/2})\zeta\left(\frac{n}{2}+1\right)\bigg\}\;.
\eea

We consider, at this point, the physical case of four dimensions.
By setting $n=4$ in (\ref{76}) we obtain
\be
G_{0}^{\rm scalar}(0,0)=\frac{1}{12}\;.
\ee
Now, we notice the following
relation
\be
(1-2^{2-n/2})\zeta\left(\frac{n}{2}-1\right)=\eta\left(\frac{n}{2}-1\right)\;,
\ee
where $\eta(x)$ is the Dirichlet eta function. In the particular case of four
dimensions we have that
\be
\lim_{n\rightarrow 4}(1-2^{2-n/2})
\zeta\left(\frac{n}{2}-1\right)=\eta(1)=\ln
2\;.
\ee
By using the last remark we obtain the values of the functions
$G_{i}(n,y)$ in four dimensions
\bea
G_{1}^{\rm scalar}(0,0)
&=&-\left(\frac{1}{6}-\xi\right)\frac{1}{\pi}\ln 2\;,
\\[10pt]
G_{2}^{\rm scalar}(0,0)
&=&\frac{1}{6\pi}\ln 2 -\frac{27}{32 \pi^{3}}\zeta(3)\;,
\\[10pt]
G_{4}^{\rm scalar}(0,0)
&=&\frac{17}{96\pi}\ln 2-\frac{63}{64\pi^{3}}\zeta(3)\;,
\\[10pt]
G_{5}^{\rm scalar}(0,0)
&=&-\frac{1}{64\pi}\ln 2+\frac{27}{128\pi^{3}}\zeta(3)\;.
\eea

\subsubsection{Spinor Fields}

For spinor fields the expressions
for the non-vanishing $G_{i}^{\rm spinor}$ in the limit $x\to 0$ are
\bea\label{72}
G_{0}^{\rm spinor}(0,y)&=&
2^{[n/2]}\pi^{-n/2}\textrm{Li}_{n/2}(e^{-\pi y})
\\[10pt]
G_{1}^{\rm spinor}(0,y)&=&
-\frac{2^{[n/2]}}{12}\frac{1}{\pi^{n/2-1}}
\textrm{Li}_{\frac{n}{2}-1}(e^{-\pi y})\;,
\\[10pt]
G_{2}^{\rm spinor}(0,y)
&=&
-\frac{2^{[n/2]}}{48\pi^{n/2+1}}\bigg\{2\pi^{3}y
\textrm{Li}_{\frac{n}{2}-2}(e^{-\pi
y})
+(n+4)\pi^{2}\textrm{Li}_{\frac{n}{2}-1}(e^{-\pi y})
\nonumber\\[5pt]
&-&18\pi y\textrm{Li}_{\frac{n}{2}}(e^{-\pi y})
-9(n+2)\textrm{Li}_{\frac{n}{2}+1}(e^{-\pi y})\bigg\}\;,
\\[10pt]
G_{4}^{\rm spinor}(0,y)
&=&\frac{2^{[n/2]}}{384\pi^{n/2+1}}
\bigg\{-16\pi^{3}y\textrm{Li}_{\frac{n}{2}-2}(e^{
-\pi y})
-4\pi^{2}(2n+12-3y^{2})\textrm{Li}_{\frac{n}{2}-1}(e^{-\pi y})
\nonumber\\[5pt]
&+&12(n+12)\pi y\textrm{Li}_{\frac{n}{2}}(e^{-\pi y})
+3(n+2)(n+24)\textrm{Li}_{\frac{n}{2}+1}(e^{-\pi y})\bigg\}\;,
\\[10pt]
G_{5}^{\rm spinor}(0,y)&=&
-2^{[n/2]}\frac{3}{256\pi^{n/2+1}}
\bigg\{4\pi^{2}(4+y^{2})\textrm{Li}_{\frac{n}{2}-1}(e^{-\pi y})
+4n\pi y\textrm{Li}_{\frac{n}{2}}(e^{-\pi y})
\nonumber\\[5pt]
&+&n(n+2)\textrm{Li}_{\frac{n}{2}+1}(e^{-\pi y})\bigg\}\;.
\eea

In the particular case of $n=4$ the above results read
\bea
G_{0}^{\rm spinor}(0,y)&=&
\frac{4}{\pi^{2}}\textrm{Li}_{2}(e^{-\pi
y})\;,
\\[10pt]
G_{1}^{\rm spinor}(0,y)&=&
\frac{1}{3\pi}\ln(1-e^{-\pi y})\;,
\label{565xxx}
\\[10pt]
G_{2}^{\rm spinor}(0,y)
&=&-\frac{1}{12\pi^{3}}\bigg\{\frac{2\pi^{3}y
e^{-\pi y}}{1-e^{-\pi y}}
-8\pi^{2}\ln(1-e^{-\pi y})
\nonumber\\[5pt]
&-&18\pi y\textrm{Li}_{2}(e^{-\pi y})
-54\textrm{Li}_{3}(e^{-\pi y})\bigg\}\;,
\\[10pt]
G_{4}^{\rm spinor}(0,y)&=&
-\frac{1}{96\pi^{3}}\bigg\{\frac{16\pi^{3}y e^{-\pi y}}{1-e^{-\pi y}}
-4\pi^{2}(20-3y^{2})\ln(1-e^{-\pi y})
\nonumber\\[5pt]
&-&192\pi y\textrm{Li}_{2}(e^{-\pi y})
-504\textrm{Li}_{3}(e^{-\pi y})\bigg\}\;,
\\[10pt]
G_{5}^{\rm spinor}(0,y)&=&\frac{3}{16\pi^{3}}
\bigg\{\pi^{2}(4+y^{2})\ln(1-e^{-\pi y})
-4\pi y\textrm{Li}_{2}(e^{-\pi y})
-6\textrm{Li}_{3}(e^{-\pi y})\bigg\}\;.
\nonumber\\
&&
\label{568xxx}
\eea


In the case of spinor fields, for $n>4$, there is a well defined limit as
$y\rightarrow 0$.
In fact, by taking the massless limit,
$y\rightarrow 0$, of the expression (\ref{72}) and noticing
that
\be\label{8}
\textrm{Li}_{n}(1)=\zeta(n)\;,
\ee
we obtain
\be
\label{803}
G_{0}^{\rm spinor}(0,0)
=\frac{2^{[n/2]}}{\pi^{n/2}}
\zeta\left(\frac{n}{2}\right)\;.
\ee
Analogously, in the limit as $y\rightarrow 0$ the result for
the remaining $G_{i}^{\rm spinor}$ can be written as follows
\bea
G_{1}^{\rm spinor}(0,0)
&=&-\frac{2^{[n/2]}}{12}\pi^{1-n/2}\zeta\left(\frac{n}{2}-1\right)\;,
\\[10pt]
G_{2}^{\rm spinor}(0,0)
&=&-\frac{2^{[n/2]}}{48\pi^{n/2+1}}
\bigg\{(n+4)\pi^{2}\zeta\left(\frac{n}{2}-1\right)
-9(n+2)\zeta\left(\frac{n}{2}+1\right)\bigg\}\;,
\nonumber\\
\eea
\bea
G_{4}^{\rm spinor}(0,0)
&=&\frac{2^{[n/2]}}{384\pi^{n/2+1}}
\bigg\{-4\pi^{2}(2n+12)\zeta\left(\frac{n}{2}-1\right)
\nonumber\\[5pt]
&&
+3(n+2)(n+24)\zeta\left(\frac{n}{2}+1\right)\bigg\}\;,
\\[10pt]
G_{5}^{\rm spinor}(0,0)
&=&-2^{[n/2]}\frac{3}{256\pi^{n/2+1}}
\bigg\{16\pi^{2}\zeta\left(\frac{n}{2}-1\right)
+n(n+2)\zeta\left(\frac{n}{2}+1\right)\bigg\}\;.
\nonumber\\
\eea

We turn our attention, now, to the physical case of $n=4$. From the expression
in
(\ref{803}) we obtain the following result
\be
G_{0}^{\rm spinor}(0,0)=\frac{2}{3}\;.
\ee
It is evident, from the expressions in
(\ref{565xxx})-(\ref{568xxx}), that the functions
\\
$G_{i}^{\rm spinor}(0,y)$ in
four dimensions represent a special case since there is an infrared
divergence as $m\rightarrow 0$ (or $y\rightarrow 0$). This means that there
is no well-defined value for the massless limit $y\rightarrow 0$. Instead,
we find a logarithmic divergence, $\log (\pi y)$. In order to
analyze this case we set $n=4$ from the beginning in the expressions
for finite $y$, and then we examine the asymptotics as $y\rightarrow 0$.
By using the equations (\ref{565xxx})-(\ref{568xxx}) we obtain
\bea
G_{1}^{\rm spinor}(0,y)
&=&\frac{1}{3\pi}\log(\pi y)+O(y)\;,
\label{5120xxx}
\\[10pt]
G_{2}^{\rm spinor}(0,y)
&=&\frac{2}{3\pi}\log(\pi y)
-\frac{1}{6\pi}
+\frac{9}{2\pi^{3}}\zeta(3)+O(y)\;,
\\[10pt]
G_{4}^{\rm spinor}(0,y)
&=&\frac{5}{6\pi}\log(\pi y)
-\frac{1}{6\pi}
+\frac{21}{4\pi^{3}}\zeta(3)+O(y)\;,
\\[10pt]
G_{5}^{\rm spinor}(0,y)
&=&\frac{3}{4\pi}\log(\pi y)
+\frac{9}{8\pi^{3}}\zeta(3)+O(y)\;.
\label{5123xxx}
\eea

Notice that, in four dimensions the functions
$G_i^{\rm spinor}(x,y)$ are singular at
the point $x=y=0$. In particular, the limits $x\to 0$
and
$y\to 0$ are not commutative, that is, the limits
as $x\to 0$
of the eqs.
(\ref{564xxx})-(\ref{573xxx})
(obtained as $y\to 0$ for a finite $x$)
 are different from the eqs.
(\ref{5120xxx})-(\ref{5123xxx})
(obtained as $y\to 0$ for $x=0$).

\section{Concluding Remarks}

In this paper we have continued the study of the heat kernel and the effective
action for complex (scalar and spinor) quantum fields in a strong constant
electromagnetic field and a gravitational field initiated in
\cite{avramidi08b}. We study here an
{\it essentially non-perturbative regime} when
the electromagnetic field is so strong that one has to take into account all its
orders. In this situation the standard asymptotic expansion of the heat kernel
does not apply since the electromagnetic field can not be treated as a
perturbation. In \cite{avramidi08b} we established the existence of a new
non-perturbative asymptotic expansion of the heat kernel and computed
explicitly the first three coefficients of this expansion.

In the present paper
we computed the first two coefficients (of zero and the first order in the
Riemann curvature) explicitly in $n$-dimensions by using the spectral
decomposition of the electromagnetic field tensor. We applied this result for
the calculation of the effective action in the physical pseudo-Euclidean
(Lorentzian) case and computed explicitly the imaginary part of the effective
action both in the general case and in the cases of physical interest. We
also computed the asymptotics of the obtained results for supercritical
electric fields.

We have discovered a {\it new infrared divergence} in the
imaginary part of the effective action for massless spinor fields
in four dimensions (or supercritical electric field), which is
induced purely by the gravitational corrections. This means physically that
the creation of massless spinor particles
(or massive particles in supercritical electric field)
is magnified substantially by the
presence of the gravitational field. Further analysis shows that a
similar effect occurs for any massless fields (also scalar fields) in the
second order in the Riemann curvature. This effect could have important
consequences for theories with spontaneous symmetry breakdown when the mass of
charged particles is generated by a Higgs field. Such theories would exhibit a
significant amount of created particles (in the massless limit an infinite
amount) at the phase transition point when the symmetry is restored and the
massive charged particles become massless.
That is why this seems to be an interesting {\it new physical
effect} that deserves further investigation.


\end{document}